\documentclass[12pt]{article}

\usepackage{amsmath,amssymb,graphicx,color}

\usepackage{authblk}

\usepackage[dvipsnames]{xcolor}
\usepackage{listings,enumitem,calc,hyperref,caption,mmacells,array,booktabs} 
\allowdisplaybreaks

\usepackage{longtable} 

\usepackage[utf8]{inputenc}

\usepackage[T1]{fontenc}
\usepackage{lmodern} \rmfamily
\DeclareFontShape{T1}{lmr}{b}{sc}{<->ssub*cmr/bx/sc}{}
\DeclareFontShape{T1}{lmr}{bx}{sc}{<->ssub*cmr/bx/sc}{}

\definecolor{comment}{rgb}{0,0.3,0}
\makeatletter
\def\@maketitle{%
  \newpage
  \null
  \vskip 2em%
  \begin{center}%
  \let \footnote \thanks
    {\LARGE \@title \par}%
    \vskip 1.5em%
    {\large \@author \par}%
    \vskip 1em%
    {\large \@date}%
  \end{center}%
  \par
  \vskip 1em}
\makeatother

\makeatletter
\def\blfootnote{\gdef\@thefnmark{}\@footnotetext}
\makeatother

\title{\vspace{-10mm}\textbf{\textsc{H1jet}, a fast program to compute transverse momentum distributions}\blfootnote{\textsc{H1jet} can be obtained from ref.~\cite{h1jet}.}}

\author{Alexander Lind} 
\author{Andrea Banfi} 
\affil{Department of Physics and Astronomy, University of Sussex, BN1 9QH Brighton, UK}

\date{\small E-mail: \href{mailto:a.lind@sussex.ac.uk}{a.lind@sussex.ac.uk}, \href{mailto:a.banfi@sussex.ac.uk}{a.banfi@sussex.ac.uk}}

\begin{document}

\maketitle

\abstract{We present \textsc{H1jet}, a fast code that computes the
  total cross section and differential distribution in the transverse
  momentum of a colour singlet. In its current version, the program
  implements only leading-order $2\to 1$ and $2\to 2$ processes, but
  could be extended to higher orders. We discuss the processes
  implemented in \textsc{H1jet}, give detailed instructions on how to
  implement new processes, and perform comparisons to existing codes. This
  tool, mainly designed for theorists, can be fruitfully used to
  assess deviations of selected new physics models from the Standard
  Model behaviour, as well as to quickly obtain distributions of
  relevance for Standard Model phenomenology.} 

\tableofcontents 

\newpage 

\section{Introduction}
\label{sec:intro}

After the discovery of the Higgs boson in
2012~\cite{Aad:2012tfa,Chatrchyan:2012ufa}, one of the most urgent tasks of
the Large Hadron Collider (LHC) is the characterisation of the Higgs
sector, in order to shed light on the exact mechanism for electroweak
symmetry breaking. In particular, Higgs production data, either in the
form of signal strengths~\cite{Aad:2019mbh,Sirunyan:2018koj} or cross
section measurements~\cite{Aaboud:2018ezd,Sirunyan:2018sgc}, offer
powerful constraints on Higgs anomalous couplings.

However, it has been pointed out that inclusive Higgs production
through gluon fusion, the one with the largest rate, is not able to
discriminate effectively between the Standard Model (SM) and another
theory giving the same effective coupling between the gluons and the
Higgs. In fact, top quarks running in loops give a dimension-6
effective interaction between the incoming gluons and the Higgs, with
exact top-mass effects giving tiny corrections. Therefore, in many
theories, the strength of dimension-6 contact gluon-gluon-Higgs
interactions can conspire with an anomalous top Higgs Yukawa coupling
to give exactly the same cross section for Higgs production as the
SM~\cite{Azatov:2013xha,Grojean:2013nya,Banfi:2013yoa}.

There are essentially two ways of solving this problem. One is to put
a direct constraint on the top Yukawa coupling by the
observation of top quarks in association with the
Higgs~\cite{Aad:2020ivc,Sirunyan:2020sum,Maltoni:2016yxb}. The other is
to break the top loop by looking e.g.\ at Higgs production at large
transverse momentum, where the Higgs recoils against a hard
jet~\cite{Azatov:2013xha,Grojean:2013nya,Banfi:2013yoa}. Both are
indirect probes of new physics effects. The latter is more difficult
experimentally, in that it relies on appreciating small deviations from
the SM in the shape of the Higgs transverse momentum distribution, in
a region where the phase space closes. It nevertheless can give a
direct access to new physics coupling the gluons to the Higgs through loops.

Higgs sector aside, production of colour singlets at high transverse
momentum is commonly used as a probe of new physics. A relevant
example is the production of monojets, which can recoil
either against dark matter, or against a SM particle decaying into
invisible particles (see e.g.\ refs.~\cite{Sirunyan:2017hci,Aaboud:2019yqu}).

Theoretical predictions for the transverse momentum distribution of a
colour singlet, both in the SM and beyond, can be currently obtained
with Monte Carlo programs, such as
\textsc{\linebreak MadGraph5\_aMC@NLO}~\cite{Alwall:2014hca} or
\textsc{SusHi}~\cite{Harlander:2012pb,Harlander:2016hcx}. These codes,
although general, have the drawback of being quite slow. Also,
interference terms between new physics and the SM, which
carry information on the strength of new interactions, are difficult
to extract from Monte Carlo event generators because they are not positive
definite. The aim of this paper is to describe a method to obtain the
transverse momentum spectrum of a colour singlet in a second or less,
and its concrete implementation in the program \textsc{H1jet}. This program
makes it possible to predict the effects of several models in a short
amount of time. This in turn opens the way to devising more refined
cut-based search strategies only for the models showing the largest
deviations with respect to the SM.

More precisely, \textsc{H1jet} predicts the transverse momentum distribution of
a colour singlet fully integrated over rapidity, and completely
inclusive with respect to all coloured particles, i.e.\ the recoiling
jets. Such an approximation is not too unrealistic, because the higher the
transverse momentum, the more the colour singlet is central, and the
more its decay products will be likely to pass the detector acceptance
cuts.

The program is based on elementary analytic manipulations on the expression
for the transverse momentum distribution. These make it possible to
write the spectrum as a one dimensional integral, whose integrand is
the product of an amplitude squared, which can be provided by ourselves or by the user, and a parton luminosity, which we extract from an external
program. The relevant amplitudes can be
either hard coded, or computed automatically and embedded in the program
via a simple user interface.

\textsc{H1jet} already comes with a number of hard-coded processes and models. 
The main process is 
\begin{equation}
    pp \to H + \mathrm{jet} \,, 
\end{equation}
where the initial state consists of gluons and light quarks, and Higgs
production procedes via quark loops. This process can be calculated in
\textsc{H1jet} for different physics models including the SM, a CP-odd
Higgs, a simplified SUSY model, and composite Higgs models with a
single or multiple top-partners. In addition, the
$b\bar{b} \to H + \mathrm{jet}$ and $pp \to Z + \mathrm{jet}$
processes for the SM are implemented. Moreover, \textsc{H1jet} is very
flexible and can be easily interfaced to use a custom user-specified
process.

The paper is organised as follows. In section~\ref{sec:method} we
briefly describe the method underlying \textsc{H1jet}. In
section~\ref{sec:manual} we describe in detail how \textsc{H1jet}
works in practice. In particular, we show how it can be installed and
run, and present the features currently implemented. In
section~\ref{sec:benchmarking} we present a detailed comparison with
the existing program \textsc{SusHi} for Higgs production both in the
SM and beyond. In section~\ref{sec:newprocs} we explain how a user can
implement a model of new physics inside \textsc{H1jet}. We choose
axion-like-particle (ALP) production, giving rise to a monojet. We
then describe how to obtain ALP transverse momentum distributions from
the generation of the Feynman rules with \textsc{FeynRules}, to the
calculation of the amplitude with \textsc{FeynCalc} and its subsequent
interface with \textsc{H1jet} to obtain the ALP transverse momentum
spectrum. Last, section~\ref{sec:the-end} presents our conclusions.

\section{The Method}
\label{sec:method}
Before explaining how the \textsc{H1jet} method works, it is instructive to
consider first how to compute the Born cross section for producing a
colour singlet $X$, e.g.\ a Higgs, of mass $m_X$. This will also allow
us to set the notation for the rest of the paper. We consider the
$2\to 1$ process $p_1 p_2 \to X$, where $p_1$ and $p_2$ are the two incoming
partons and $p_X$ is the momentum of the considered colour
singlet. From momentum conservation we have 
\begin{equation}
  \label{eq:mom-conservation}
  \hat s = (p_1+p_2)^2=m_X^2\,.
\end{equation}
There can be various partonic subprocesses that contribute to the
production of the particle $X$. Let us denote with $M_{ij}$ the
amplitude for the subprocess $ij\to X$ (e.g. $gg \to H$), with
$i,j=g,q_f,\bar q_{\bar f}$, where $f,\bar f$ denotes quark or
antiquark flavours. The Born partonic cross section for each
subprocess is
\begin{equation}
  \label{eq:Born-xsct}
  \hat\sigma_{ij}=\frac{\pi}{m_X^2} |M_{ij}(m_X^2)|^2 \delta\left(\hat s-m_X^2\right)\,.
\end{equation}
The corresponding hadronic cross section is given by
\begin{equation}
  \label{eq:Born-had}
  \begin{split}
  \sigma& =\frac{\pi}{m_X^2}\sum_{i,j} \int_0^1 dx_1\, f_{i/p}(x_1,\mu_F)\int_0^1 dx_2 \, f_{i/p}(x_1,\mu_F)\sum_{i,j}
 |M_{ij}(m_X^2)|^2 \delta\left(x_1 x_2 s-m_X^2\right)\\
   & = \frac{\pi}{m_X^4} \sum_{i,j} |M_{ij}(m_X^2)|^2 \mathcal{L}_{ij}\left(\frac{m_X^2}{s},\mu_F\right)
  \,,
    \end{split}
\end{equation}
where $\mathcal{L}_{ij}\left(\tau,\mu_F\right)$ is the partonic luminosity
\begin{equation}
  \label{eq:hoppet-lumi}
  \mathcal{L}_{ij}\left(\tau,\mu_F\right) = \tau \int_\tau^1 \frac{dx}{x}\, f_{i/p}(x,\mu_F)\, f_{j/p}\left(\frac{\tau}{x},\mu_F\right)\,.
\end{equation}
If we are able to obtain the luminosity
$\mathcal{L}_{ij}\left(m_X^2/s,\mu_F\right)$, we are then able to
obtain a numerical prediction for the cross section through a simple
multiplication. There are indeed numerical tools that are able to
compute, tabulate and interpolate luminosities with incredible
efficiency, for instance the program
\textsc{HOPPET}~\cite{Salam:2008qg}. Through an interface with \textsc{HOPPET}, we are
able to compute the Born cross section given the amplitudes
$M_{ij}$. This procedure is the same adopted in the program
\textsc{JetVHeto}~\cite{Banfi:2015pju}, that computes cross sections for colour
singlets with a veto on additional jets.

A similar strategy can be devised to obtain a fast calculation of
distributions in the transverse momentum of particle $X$. A non-zero
transverse momentum for $X$ is obtained via a generic $2\to 2$
partonic process $p_1 p_2 \to p_3 X$, where $p_1$, $p_2$, and $p_3$ are massless
partons, and $p_X$ is the momentum of the colour singlet $X$. We
wish to compute $d\sigma/dp_T$, where $p_T$ is the transverse momentum
of $p_X$ with respect to the beam axis. At Born level only, $p_T$ is
also the transverse momentum of the recoiling jet originated by
$p_3$. The partonic subprocesses contributing to $d\sigma/dp_T$ are
$gg\to g X$, $q_f \bar q_{\bar f} \to g X$, $q_f g\to q_f X$,
$gq_f \to q_fX$. The corresponding amplitudes $M_{ij}$ (with
$i,j=g,q_f,\bar q_{\bar f}$) are functions of the three Mandelstam
invariants
\begin{equation}
  \label{eq:Mandelstam}
  \begin{split}
  \hat s & = (p_1+p_2)^2 = (p_3+p_X)^2 \,,\\
  \hat t & = (p_1-p_3)^2 = (p_2-p_X)^2 \,,\\
  \hat u & = (p_2-p_3)^2 = (p_1-p_X)^2 \,.
  \end{split}
\end{equation}
Without loss of generality, in the centre-of-mass frame of the
partonic collision, we can parameterise momenta as follows
\begin{equation}
  \label{eq:momenta}
  \begin{split}
  p_1&=\frac{\sqrt {\hat s}}{2}(1,0,0,1)\,,\qquad p_3 = p_T(\cosh\eta,1,0,\sinh\eta)\,,\\
  p_2&=\frac{\sqrt {\hat s}}{2}(1,0,0,-1)\,,\qquad p_X = \left ( \sqrt{m_X^2+p^2_T\cosh^2\eta},-p_T,0,-p_T \sinh\eta \right ) \,,
  \end{split}
\end{equation}
where $\eta$ is the rapidity of parton $p_3$ in the
centre-of-mass frame. The {\em partonic} $p_T$ spectrum for the
process initiated by partons $ij$ is given by
\begin{equation}
  \label{eq:partonic-pt}
  \frac{d\hat \sigma_{ij}}{dp_T} = \frac{p_T}{16\pi}\int d\eta \frac{|M_{ij}(\hat s,\hat t,\hat u)|^2}{E_X \hat s} \delta\left(\sqrt{\hat s} - p_T\cosh\eta-\sqrt{m_X^2+p^2_T\cosh^2\eta}\right)\,, 
\end{equation}
where $E_X=\sqrt{m_X^2+p^2_T\cosh^2\eta}$ is the energy of the colour-singlet
particle $p_X$.  The above equation selects two values of $\eta$, as
follows
\begin{equation}
\label{eq:eta-values}
  \eta = \ln\left(\hat x_M \pm \sqrt{\hat x_M^2-1}\right)\,,\qquad \hat x_M\equiv\frac{\hat s-m_X^2}{2 p_T \sqrt{\hat s} }\,.
\end{equation}
The corresponding hadronic cross section reads
\begin{equation}
\label{eq:hadronic-pt}
  \frac{d\sigma}{dp_T} = \sum_{i,j} \int_0^1 dx_1\, f_{i/p}(x_1,\mu_F)\int_0^1 dx_2 \, f_{i/p}(x_1,\mu_F) \left[\frac{d\hat \sigma_{ij}}{dp_T}\, \Theta\left(\hat s-p_T-\sqrt{m_X^2+p^2_T}\right)  \right]_{\hat s=x_1x_2 s}\,.
\end{equation}
Since eq.~\eqref{eq:eta-values} gives two monotonic functions of $\hat s$ for $s>p_T+\sqrt{m_X^2+p^2_T}$, varying $\hat s$ in the allowed range spans all possible values of $\eta$ in the range $-\eta_M<\eta<\eta_M$ with
\begin{equation}
\label{eq:eta-range}
\eta_M \equiv \ln \left ( x_M + \sqrt{x_M^2-1} \right ) \,, \qquad x_M\equiv\frac{s-m_X^2}{2 p_T \sqrt{s}} \,.
\end{equation}
This allows us to perform the $\eta$ integration last, and obtain, after some
manipulations,
\begin{equation}
\label{eq:hadronic-pt-lumi}
  \frac{d\sigma}{dp_T} = \frac{p_T}{8 \pi}\int_{-\eta_M}^{\eta_M}\!d\eta\, \sum_{i,j}  \left[\frac{M^2_{ij}\left(\hat s,\hat t=-p_T e^{-\eta}\sqrt {\hat s}  ,\hat u =- p_T e^{\eta}\sqrt{\hat s}\right)}{E_X \hat s^{3/2}}\mathcal{L}_{ij}\left(\frac{\hat s}{s},\mu_F\right) \right]_{\hat s=\left(p_T\cosh\eta+\sqrt{m_X^2+p^2_T\cosh^2\eta}\right)^2}\,,
\end{equation}
where again $\mathcal{L}_{ij}$ is the partonic luminosity for the $ij$
incoming channel as defined in eq.~\eqref{eq:hoppet-lumi}. If we are
able to obtain the partonic luminosity $\mathcal{L}_{ij}$, say, from
\textsc{HOPPET}, we can obtain the transverse momentum spectrum with a
one-dimensional integration, which can be performed extremely quickly
with a Gaussian numerical integrator.

Summarising, by interfacing \textsc{HOPPET} with a code that provides
amplitudes for $2\to 1$ and $2\to 2$ partonic subprocesses producing a
colour singlet $X$, we are able to perform fast computations of total
cross sections and transverse momentum spectra for $X$. In the
following sections we describe our implementation of the method for
Born processes. Note that, if one were able to perform the analytic
integration over the phase space of final-state partons, the method
can also be applied to higher-order cross sections and differential
spectra.

\section{User's Manual}
\label{sec:manual} 

This section describes the most important technical details of
\textsc{H1jet}, including its installation and usage.

\subsection{Installation} 
The source code of \textsc{H1jet} can be obtained from
ref.~\cite{h1jet}. The source code consists of a main directory
\texttt{\textsc{H1jet}} with the following subdirectories:
\begin{description}[labelindent=1cm, labelwidth =\widthof{\bfseries9999}, leftmargin = !]
	\item[\texttt{bin} :] contains the executable program \texttt{h1jet} after compilation, as well as the Python 3 helper scripts \texttt{PlotH1jet.py} and \texttt{DressUserAmpCode.py}. 
	\item[\texttt{src} :] source files. 
\end{description}
The \texttt{README.md} file contains information on installation and usage. \\

\noindent
In the main directory, the user needs to run the configure script: 
\begin{lstlisting}
	./configure [options] 
\end{lstlisting}
It will attempt to find a Fortran compiler (\texttt{gfortran} or
\texttt{ifort}), as well as the dependencies on the user's machine. A
specific compiler and/or compiler flags can be selected with the
options \texttt{./configure FC=<compiler>} and \texttt{./configure
  FFLAGS=<flags>}. \textsc{H1jet} has a number of external
dependencies which it must be linked to:
\begin{itemize}
	\item \textsc{LHAPDF}~\cite{Buckley:2014ana}: Provides the PDF sets for \textsc{H1jet}. 
	\item \textsc{HOPPET}~\cite{Salam:2008qg}: For QCD DGLAP
          evolution of PDFs and numerical integrations.
	\item \textsc{CHAPLIN}~\cite{Buehler:2011ev}: For complex
          harmonic polylogarithms used to represent scalar integrals
          in loop-induced processes.
\end{itemize} 
For the \textsc{CHAPLIN} library, it may be necessary to explicitly state the path to the library files with:
\begin{lstlisting}
	./configure LDFLAGS=-L/path/to/chaplin/lib 
\end{lstlisting}
To compile with a custom user interface: 
\begin{lstlisting}
	./configure USERFILE=/path/to/custom/user_interface.f90  
\end{lstlisting}
See Section \ref{sec:newprocs} below for the implementation of custom user-specified amplitudes. \\ 
To install in a specific location: 
\begin{lstlisting}
	./configure --prefix=/path/to/installation 
\end{lstlisting}
The default installation path is the main \texttt{H1jet}-directory. \\ 
The configure script will generate the \texttt{Makefile}. \\ 
To compile \textsc{H1jet} with the generated \texttt{Makefile}, run: 
\begin{lstlisting}
	make [options] 
\end{lstlisting}
This command takes the following options: \texttt{make clean} will delete all module and object files; \texttt{make distclean} will delete all module and object files as well as the executable \texttt{h1jet}. \\ 

After compilation, the \texttt{bin}-directory can then be added to the
user's \texttt{PATH} environment variable. Alternatively, if the user
has specified an installation directory with the \texttt{-{}-prefix}
option, the executable can be installed with:
\begin{lstlisting}
	make install  
\end{lstlisting}
The executable \texttt{h1jet} can then be found in the \texttt{bin}-directory at the path specified by \texttt{-{}-prefix}. 

\subsection{Usage} 
After compilation, \textsc{H1jet} can be run from the \texttt{bin}-directory with: 
\begin{lstlisting}
	./bin/h1jet [options]  
\end{lstlisting}
\textsc{H1jet} will print out a brief summary of the settings
and parameters used, as well as the Born cross section
$\sigma_0$, followed by a five-column table. The first three entries of each row specifies the lower end, the midpoint, and the upper end of each $p_T$ bin. The fourth entry is $d\sigma/dp_T$ evaluated at the midpoint of the corresponding $p_T$ bin. The fifth entry is the integrated cross section $\sigma(p_T)$ with a lower bound in $p_T$ corresponding to the lower end of the given $p_T$ bin. We remark that the fundamental object we compute is $d\sigma/dp_T$. The integrated cross section $\sigma(p_T)$ is obtained by summing $d\sigma/dp_T$ over the appropriate range and multiplying by the bin width. Therefore, this procedure gives a reliable estimate of $\sigma(p_T)$ only if the binning is fine enough.

      \noindent      
The following standard UNIX options are available: 
\begin{description}[labelindent=1cm, labelwidth =\widthof{\bfseries9999999999999999999999}, leftmargin = !] 
	\item[\texttt{-h, -{}-help}] Display the help message along with all possible options. 
	\item[\texttt{-v, -{}-version}] Display the version of the installed \textsc{H1jet}. 
\end{description}
\textsc{H1jet} will display the requested information and then terminate. 

\noindent
The output can be directed to a file with the option: 
\begin{description}[labelindent=1cm, labelwidth =\widthof{\bfseries9999999999999999999999}, leftmargin = !] 
	\item[\texttt{-o, -{}-out <file>}] Direct the output to \texttt{<file>}.\\ Default: standard output.
\end{description}
The physics process can be selected with: 
\begin{description}[labelindent=1cm, labelwidth =\widthof{\bfseries9999999999999999999999}, leftmargin = !] 
	\item[\texttt{-{}-proc <arg>}] Specify the process. Arguments: \vspace{-2mm} 
	\begin{description}[labelwidth =\widthof{\bfseries99999}, leftmargin = !] 
		\item[\texttt{H}] $pp/p\bar{p} \rightarrow H + \text{jet}$ (default). 
		\item[\texttt{bbH}] $b\bar{b} \rightarrow H + \text{jet}$. 
		\item[\texttt{Z}] $pp/p\bar{p} \rightarrow Z + \text{jet}$. 
		\item[\texttt{user}] User specified process. See Section \ref{sec:newprocs} below for details on the implementation. 
	\end{description}
\end{description}
Depending on the process selected, there exists different relevant options. 

\subsubsection{General Options}
The options listed here apply to all processes. 
\begin{description}[labelindent=1cm, labelwidth =\widthof{\bfseries9999999999999999999999}, leftmargin = !] 
	\item[\texttt{-{}-collider <arg>}] Specify the collider type. \\ Arguments: \texttt{pp} (default), \texttt{ppbar}. 
	\item[\texttt{-{}-roots <value>}] Centre-of-mass energy, $\sqrt{s}$ [GeV]. \\ Default: 13000 GeV. 
	\item[\texttt{-{}-pdf\_name <arg>}] Specify the PDF set name from \textsc{LHAPDF}. \\ The specified PDF set must be available in the local installation of \textsc{LHAPDF}. \\ Default: \texttt{MSTW2008nlo68cl}. 
	\item[\texttt{-{}-pdf\_mem <value>}] Integer value specifying the PDF member. \\ Default: 0. 
	\item[\texttt{-{}-scale\_strategy <arg>}] Set the scale strategy, i.e. the dynamic $\mu = \mu_R = \mu_F$ value. \\ Arguments: \vspace{-2mm} 
	\begin{description}[labelwidth =\widthof{\bfseries99999}, leftmargin = !] 
		\item[\texttt{M}] $\mu = M$. 
		\item[\texttt{HT}] $\mu = p_T + \sqrt{p_T^2 + M^2}$ (default). 
		\item[\texttt{MT}] $\mu = \sqrt{p_T^2 + M^2}$. 
	\end{description} \vspace{-1mm} 
	The mass $M$ is given by option \texttt{-{}-mH} for processes \texttt{H} and \texttt{bbH}, option \texttt{-{}-mZ} for process \texttt{Z}, and option \texttt{-{}-mass} for process \texttt{user}. 
      \item[\texttt{-{}-xmur <value>}] Additional factor $x_R$ for the
        renormalisation scale, i.e.\ $\mu_R=x_R\,\mu$, where $\mu$ is
        determined by the choice from \texttt{-{}-scale\_strategy}. \\
        Default: 0.5.
      \item[\texttt{-{}-xmuf <value>}] Additional factor $x_F$ for the
        factorisation scale, i.e.\ $\mu_F=x_F\,\mu$, where $\mu$ is
        determined by the choice from \texttt{-{}-scale\_strategy}. \\
        Default: 0.5.
      \item[\texttt{-{}-nbins <value>}] Number of histogram bins in the
        output of the transverse momentum distribution. \\ Default:
        400.
      \item[\texttt{-{}-log}] Enables logarithmic $x$-axis of the
        histogram, i.e.\ logarithmic bins in $p_{T}$. The option
        \texttt{-{}-ptmin} must be set to a non-zero value in order to
        use this option, otherwise the program will quit with an
        error.
	\item[\texttt{-{}-ptmin <value>}] Minimum $p_{T}$ value [GeV]. \\ Default: 0 GeV. 
	\item[\texttt{-{}-ptmax <value>}] Maximum $p_{T}$ value [GeV]. \\ Default: 4000 GeV. 
	\item[\texttt{-{}-accuracy <value>}] The desired integration accuracy. \\ Default: 0.001. 
\end{description}

\subsubsection{Relevant Options for Process: \texttt{H}}
If process \texttt{H} is selected, i.e.\ 
$	pp/p\bar{p} \rightarrow H + \text{jet}$,
then the following options are relevant: 
\begin{description}[labelindent=1cm, labelwidth =\widthof{\bfseries9999999999999999999999}, leftmargin = !] 
	\item[\texttt{-{}-mH <value>}] Higgs mass, $m_H$ [GeV]. \\ Default: $125$ GeV. 
	\item[\texttt{-{}-mW <value>}] W boson mass, $m_W$ [GeV]. \\ Default: $80.385$ GeV. 
	\item[\texttt{-{}-mZ <value>}] Z boson mass, $m_Z$ [GeV]. \\ Default: $91.1876$ GeV.
	\item[\texttt{-{}-mt <value>}] Top quark mass, $m_t$ [GeV]. \\ Default: $173.5$ GeV. 
	\item[\texttt{-{}-mb <value>}] On-shell bottom quark mass, $m_b^{\text{OS}}$ [GeV]. \\ Default: 4.65 GeV. 
	\item[\texttt{-{}-yt <value>}] Top Yukawa factor, $\kappa_t$ [GeV]. \\ Default: $1$. 
	\item[\texttt{-{}-yb <value>}] Bottom Yukawa factor, $\kappa_b$
          [GeV]. \\ Default: $1$ ($0$ for CP-odd Higgs).
	\item[\texttt{-{}-GF <value>}] Fermi coupling constant, $G_F$ [GeV$^{-2}$]. \\ Default: $0.116638 \times 10^{-4}$ GeV$^{-2}$. 
\end{description}
Note that the Yukawa couplings are given by $y_q=\kappa_q (\sqrt 2 m_q / v)$, where $\kappa_q$ are the dimensionless factors specified by the options \texttt{-{}-yt} and \texttt{-{}-yb} above, and $v/\sqrt{2}$ is the vacuum expectation value of the Higgs field. \\ 

Note also that \textsc{H1jet} uses the $G_{\mu}$ scheme for the all electroweak parameters~\cite{Georgi:1991ci}. Hence, the Higgs vacuum expectation value is given by $v = 1 / \sqrt{\sqrt{2} G_F}$, and the Weinberg angle is $\sin^2 \theta_W = 1 - m_W^2 / m_Z^2$. \\ 

\noindent
To consider a CP-odd Higgs instead, it is necessary to select the following option: 
\begin{description}[labelindent=1cm, labelwidth =\widthof{\bfseries9999999999999999999999}, leftmargin = !] 
	\item[\texttt{-{}-cpodd}] Toggle for calculation of CP-odd Higgs. 
\end{description}
The interaction between the CP-odd Higgs $H$ and a SM quark $q$ is: 
\begin{equation}
   \mathcal{L} \supset i \tilde{\kappa}_q \frac{m_q}{v} \bar{q} \gamma_5 q H\,,
\end{equation}
where the implementation in \textsc{H1jet} uses by default $\tilde{\kappa}_t = 1$ and $\tilde{\kappa}_b = 0$. Both parameters can be changed with the options \texttt{-{}-yt} and \texttt{-{}-yb}.

Here, both CP-even and CP-odd Higgs production are loop-induced
processes. The amplitudes for $2\to 1$ processes are taken from
ref.~\cite{Spira:1995rr}. For CP-even Higgs production in $2\to 2$,
the amplitudes are taken from ref.~\cite{Baur:1989cm}, and their
interface is adapted from \textsc{HERWIG 6}~\cite{Corcella:2000bw}. We have taken the CP-odd $2\to 2$
amplitudes from ref.~\cite{Grojean:2013nya}.

\paragraph{Top-partner.} \textsc{H1jet} allows the calculation of
Higgs production via loops of top partners in addition to top loops.
To include a top-partner $T$ in the quark loops, it is necessary
to set the top-partner mass $m_T$ to a non-zero value by using the
\texttt{-{}-mtp} option.

The SM top Yukawa factor can be modified by the mixing angle, 
\begin{equation}
   \kappa_t \rightarrow \kappa_t \cos^2\theta_T \,, 
\end{equation} 
where $\kappa_t$ is the Yukawa factor set by option \texttt{-{}-yt}. 

The top-partner Yukawa factor will likewise be modified 
\begin{equation}
   \kappa_T \rightarrow \kappa_T \sin^2\theta_T \,. 
\end{equation} 
with $\kappa_T$ set by \texttt{-{}-ytp}.

\noindent
The top-partner specific options are: 
\begin{description}[labelindent=1cm, labelwidth =\widthof{\bfseries9999999999999999999999}, leftmargin = !] 
	\item[\texttt{-{}-mtp <value>}] Top-partner mass, $m_{T}$ [GeV]. \\ Default: 0 GeV. 
	\item[\texttt{-{}-ytp <value>}] Top-partner Yukawa factor, $\kappa_T$. \\ Default: 1. 
	\item[\texttt{-{}-sth2 <value>}] Top-partner mixing angle, $\sin^2 \theta_T$. \\ Default: 0. 
        \end{description}
        The above is for a simplified composite Higgs model, where the
        compositeness scale $f$ is set to infinity.  The top-partner
        can also be considered in the explicit composite Higgs models
        of ref.~\cite{Banfi:2019xai}, all with finite $f$. Four
        different models are implemented, $\textbf{M1}_{\textbf{5}}$,
        $\textbf{M1}_{\textbf{14}}$, $\textbf{M4}_{\textbf{5}}$, and
        $\textbf{M4}_{\textbf{14}}$, which modify the Yukawa
        coupling factors in the following way:
\begin{longtable}{ll}
		$\textbf{M1}_{\textbf{5}}$ & $\displaystyle\begin{array}{l} 
		\kappa_b \rightarrow \kappa_b c_\varepsilon \\ 
		\kappa_t \rightarrow \kappa_t c_\varepsilon \cos^2 \theta_L \\ 
		\kappa_T \rightarrow \kappa_T c_\varepsilon \sin^2 \theta_L \\ 
		\tilde{\kappa}_b = \tilde{\kappa}_t = \tilde{\kappa}_T = 0 
	\end{array}$ \\ \midrule
		$\textbf{M1}_{\textbf{14}}$ & $\displaystyle\begin{array}{l} 
		\kappa_b \rightarrow \kappa_b \frac{2 c_\varepsilon^2 - 1}{c_\varepsilon} \\ 
		\kappa_t \rightarrow \kappa_t \frac{2 c_\varepsilon^2 - 1}{c_\varepsilon} \cos^2\theta_L  \\ 
		\kappa_T \rightarrow \kappa_T \frac{2 c_\varepsilon^2 - 1}{c_\varepsilon} \sin^2\theta_L  \\ 
		\tilde{\kappa}_b = \tilde{\kappa}_t = \tilde{\kappa}_T = 0 
	\end{array}$ \\ \midrule
		$\textbf{M4}_{\textbf{5}}$ & $\displaystyle\begin{array}{l} 
		\kappa_b \rightarrow \kappa_b c_\varepsilon \\ 
		\kappa_t \rightarrow \kappa_t c_\varepsilon \left ( \cos^2\theta_R - \frac{s_\varepsilon^2}{1 + c_\varepsilon^2} \left ( \cos^2\theta_L - \cos^2\theta_R \right ) \right ) \\ 
		\kappa_T \rightarrow \kappa_T c_\varepsilon \left ( \sin^2\theta_R - \frac{s_\varepsilon^2}{1 + c_\varepsilon^2} \left ( \sin^2\theta_L - \sin^2\theta_R \right ) \right ) \\ 
		\tilde{\kappa}_b = 0 \\
		\tilde{\kappa}_t = \frac{4 c_\varepsilon s_\varepsilon}{\sqrt{2(1 + c_\varepsilon^2)}} \text{Im}(c_1) \sin\theta_R \cos\theta_R \\ 
		\tilde{\kappa}_T = -\tilde{\kappa}_t 
	\end{array}$ \\ \midrule
		$\textbf{M4}_{\textbf{14}}$ & $\displaystyle\begin{array}{l} 
		\kappa_b \rightarrow \kappa_b \frac{2 c_\varepsilon^2 - 1}{c_\varepsilon} \\ 
		\kappa_t \rightarrow \kappa_t \left ( \frac{(2 c_\varepsilon^2 - 1)}{c_\varepsilon} \cos^2\theta_R - \frac{s_\varepsilon^2 c_\varepsilon (8 c_\varepsilon^2 - 3)}{1 - 3 c_\varepsilon^2 + 4 c_\varepsilon^4} \left ( \cos^2\theta_L - \cos^2\theta_R \right ) \right ) \\ 
		\kappa_T \rightarrow \kappa_T \left ( \frac{(2 c_\varepsilon^2 - 1)}{c_\varepsilon} \sin^2\theta_R - \frac{s_\varepsilon^2 c_\varepsilon (8 c_\varepsilon^2 - 3)}{1 - 3 c_\varepsilon^2 + 4 c_\varepsilon^4} \left ( \sin^2\theta_L - \sin^2\theta_R \right ) \right ) \\ 
		\tilde{\kappa}_b = 0 \\
		\tilde{\kappa}_t = \frac{4 s_\varepsilon (1 - 2 s_\varepsilon^2)}{\sqrt{2(1 - 3 c_\varepsilon^2 + 4 c_\varepsilon^4)}} \text{Im}(c_1) \sin\theta_R \cos\theta_R \\ 
		\tilde{\kappa}_T = -\tilde{\kappa}_t 
	\end{array}$ \\ 
\end{longtable}
where the $\tilde{\kappa}$'s are the CP-odd couplings, and, 
\begin{equation}
	s_\varepsilon = \frac{v}{f} \,, \hspace{7 mm} \text{ and } \hspace{7 mm} c_\varepsilon = \sqrt{1 - s_\varepsilon^2} \,. 
\end{equation}
For $\textbf{M1}_{\textbf{5}}$ and $\textbf{M1}_{\textbf{14}}$, the
option \texttt{-{}-sth2} sets the mixing angle $\theta_L$, while for
$\textbf{M4}_{\textbf{5}}$ and $\textbf{M4}_{\textbf{14}}$, the same
option sets the angle $\theta_R$. The reason for this is that we want
to reproduce the $f\to \infty$ limit, where
$\theta_T=\theta_L,\theta_R$ depending on the chosen model. When
needed, the angles $\theta_L$ and $\theta_R$ are derived one from the
other by using the relation
\begin{equation}
	m_t^2 \tan^2\theta_L = m_T^2 \tan^2\theta_R \,. 
\end{equation}
The composite Higgs top-partner model specific options are: 
\begin{description}[labelindent=1cm, labelwidth =\widthof{\bfseries9999999999999999999999}, leftmargin = !] 
	\item[\texttt{-{}-model <arg>}] Specify the top-partner model. Arguments: \vspace{-2mm} 
	\begin{description}[labelwidth =\widthof{\bfseries999999}, leftmargin = !] 
		\item[\texttt{M1\_5}] $\textbf{M1}_{\textbf{5}}$, with a light top-partner transforming as a $\textbf{1}_{\textbf{2/3}}$ of $\text{SO}(4)$, and the SM top-bottom doublet embedded in a $\textbf{5}$ of $\text{SO}(5)$ (default). 
		\item[\texttt{M1\_14}] $\textbf{M1}_{\textbf{14}}$, with a light top-partner transforming as a $\textbf{1}_{\textbf{2/3}}$ of $\text{SO}(4)$, and the SM top-bottom doublet embedded in a $\textbf{14}$ of $\text{SO}(5)$. 
		\item[\texttt{M4\_5}] $\textbf{M4}_{\textbf{5}}$, with a light top-partner transforming as a $\textbf{4}_{\textbf{2/3}}$ of $\text{SO}(4)$, and the SM top-bottom doublet embedded in a $\textbf{5}$ of $\text{SO}(5)$. 
		\item[\texttt{M4\_14}] $\textbf{M4}_{\textbf{5}}$, with a light top-partner transforming as a $\textbf{4}_{\textbf{2/3}}$ of $\text{SO}(4)$, and the SM top-bottom doublet embedded in a $\textbf{14}$ of $\text{SO}(5)$. 
	\end{description} \vspace{-1mm} 
	\item[\texttt{-{}-imc1 <value>}] Imaginary part of the $c_1$ coefficient, $\text{Im}(c_1)$. \\ Default: 0.  
	\item[\texttt{-f, -{}-fscale <value>}] Compositeness scale, $f$
          [GeV]. \\ If the option is not set, all couplings will be
          automatically computed in the limit $f\to\infty$. If the
          option is set but no value is provided, the program will
          return a floating point exception.
\end{description} 

\paragraph{Multiple top-partners models.}
\textsc{H1jet} makes it possible to include multiple top-partners in
the particle loops. To do that, it will be necessary to specify an input
file with the masses and Yukawa coupling factors for each particle
running in the loop, including SM quarks. This can be done with the
following option:
\begin{description}[labelindent=1cm, labelwidth =\widthof{\bfseries9999999999999999999999}, leftmargin = !] 
\item[\texttt{-i, -{}-in <file>}] Include input file with top-partner
  masses and Yukawas. See the file \texttt{SM.dat} for the SM case,
  without any top partners.
\end{description} 
The first line of the input file should specify the number of
particles running in the loops, e.g.:
\begin{lstlisting}
	nmax: 2 
\end{lstlisting}
This should be followed by \texttt{nmax} number of lines -- one for each particle loop -- in the format of four numbers specifying the mass, $\kappa$, $\tilde{\kappa}$, and loop approximation (see later), in that order. For example, for a SM top quark: 
\begin{lstlisting}
	173.5 1.0 0.0 1 
\end{lstlisting}
with mass $173.5$ GeV, $\kappa = 1$, $\tilde{\kappa} = 0$, and the loop approximation set to $1$. \\ 

The dimensionless Yukawa coupling factors $\kappa_q$ and $\tilde{\kappa}_q$ are respectively the CP-even and CP-odd couplings for a quark $q$, with the following Lagrangian: 
\begin{equation}
	\mathcal{L} \supset -m_q \bar{q}q - \kappa_q \frac{m_q}{v} \bar{q}q H + i \tilde{\kappa}_q \frac{m_q}{v} \bar{q} \gamma_5 q H \,,
\end{equation}
where $m_q$ is the mass of the quark. \\ 

The integer value specifying the loop approximations can take the following values: 
\begin{description}[labelindent=1cm, labelwidth =\widthof{\bfseries9999}, leftmargin = !] 
	\item[$0$] Small mass limit for fermions. 
	\item[$1$] Full mass effects for fermions. 
	\item[$2$] Large mass limit for fermions. 
	\item[$3$] Full mass effects for scalars. 
	\item[$4$] Large mass limit for scalars. 
\end{description}
See Section \ref{sec:loopapprox} for more information on the loop
approximations. Note that, for implemented processes, using an input
file is the only way to change the approximation in which loops are
computed.

\paragraph{SUSY.} \textsc{H1jet} includes the simplified SUSY model with
two stops $\tilde{t}_1$ and $\tilde{t}_2$ considered in refs.~\cite{Gunion:2002zf} and~\cite{Banfi:2018pki}. To include the SUSY stops $\tilde{t}_1$ and
$\tilde{t}_2$ in the quark loops, it will be necessary to set the
first stop mass $m_{\tilde{t}_1}$ to a non-zero value by using the
\texttt{-{}-mst} option. The second stop mass is then given by
\begin{equation}
	m_{\tilde{t}_2} = \sqrt{m_{\tilde{t}_1}^2 + (\Delta m)^2} \,,
\end{equation}
where $\Delta m$ is set with the \texttt{-{}-delta} option. 

The stop Yukawa coupling factors will be given by: 
\begin{equation}
   \kappa_{\tilde{t}_1} = \frac{m_t^2}{m_{\tilde{t}_1}^2} \left [ \alpha_1 \cos^2 \theta_{\tilde{t}} + \alpha_2 \sin^2 \theta_{\tilde{t}} + 2 - \frac{(\Delta m)^2}{2 m_t^2} \sin^2\left ( 2 \theta_{\tilde{t}} \right ) \right ] \,,
\end{equation} 
\begin{equation}
   \kappa_{\tilde{t}_2} = \frac{m_t^2}{m_{\tilde{t}_2}^2} \left [ \alpha_1 \sin^2 \theta_{\tilde{t}} + \alpha_2 \cos^2 \theta_{\tilde{t}} + 2 + \frac{(\Delta m)^2}{2 m_t^2} \sin^2\left ( 2 \theta_{\tilde{t}} \right ) \right ] \,, 
\end{equation} 
where 
\begin{equation}
   \alpha_1 = \frac{m_Z^2}{m_t^2} \cos \left ( 2 \beta \right ) \left [ 1 - \frac{4}{3} \sin^2\theta_W \right ] \,, 
\end{equation} 
\begin{equation}
   \alpha_2 = \frac{4}{3} \frac{m_Z^2}{m_t^2} \cos \left ( 2 \beta \right ) \sin^2\theta_W \,. 
\end{equation} 
Note that $m_t$, $m_Z$, and $\sin^2\theta_W$ can be set with the \texttt{-{}-mt}, \texttt{-{}-mZ}, and \texttt{-{}-sinwsq} options respectively, while $\sin^2 \theta_{\tilde{t}}$ and $\tan\beta$ are SUSY specific options and can be set with the \texttt{-{}-sth2} and \texttt{-{}-tbeta} options. 

\noindent All of the SUSY specific options are: 
\begin{description}[labelindent=1cm, labelwidth =\widthof{\bfseries9999999999999999999999}, leftmargin = !] 
	\item[\texttt{-{}-mst <value>}] SUSY stop mass, $m_{\tilde{t_1}}$ [GeV]. \\ Default: 0 GeV. 
	\item[\texttt{-{}-delta <value>}] SUSY stop mass separation, $\Delta m$ [GeV]. \\ Default: 0 GeV. 
	\item[\texttt{-{}-sth2 <value>}] Stop mixing angle, $\sin^2 \theta_{\tilde{t}}$. \\ Default: 0. 
	\item[\texttt{-{}-tbeta <value>}] Ratio of VEVs of the two SUSY Higgs fields, $\tan\beta$. \\ Default: 0. 
\end{description}
Note that the top partner mass $m_{T}$ and SUSY stop mass
$m_{\tilde{t}_1}$ can not both be set non-zero at the same time via
command-line options. However, if one uses an input file, one can
explicitly specify masses, couplings and loop approximations for an
arbitrary number of fermions and scalars. This would also allow a user
to implement a specific SUSY model with more supersymmetric partners,
each with the appropriate coupling.

\subsubsection{Relevant Options for Process: \texttt{bbH}}
If process \texttt{bbH} is selected, i.e.\
$
	b\bar{b} \rightarrow H + \text{jet}
$, then the following options are relevant: 
\begin{description}[labelindent=1cm, labelwidth =\widthof{\bfseries9999999999999999999999}, leftmargin = !] 
	\item[\texttt{-{}-mH <value>}] Higgs mass, $m_H$ [GeV]. \\ Default: 125 GeV. 
	\item[\texttt{-{}-GF <value>}] Fermi coupling constant, $G_F$ [GeV$^{-2}$]. \\ Default: $0.116638 \times 10^{-4}$ GeV$^{-2}$. 
	\item[\texttt{-{}-mbmb <value>}] $\overline{\text{MS}}$ bottom quark mass, $m_b^{\overline{\text{MS}}}$ [GeV]. \\ Default: 4.18 GeV. 
\end{description}

\subsubsection{Relevant Options for Process: \texttt{Z}}
If process \texttt{Z} is selected, i.e.\
$ pp/p\bar{p} \rightarrow Z + \text{jet} $, then the following options
are relevant:
\begin{description}[labelindent=1cm, labelwidth =\widthof{\bfseries9999999999999999999999}, leftmargin = !] 
	\item[\texttt{-{}-mZ <value>}] Z boson mass, $m_Z$ [GeV]. \\ Default: 91.1876 GeV.
	\item[\texttt{-{}-mW <value>}] W boson mass, $m_W$ [GeV]. \\ Default: 80.385 GeV. 
	\item[\texttt{-{}-GF <value>}] Fermi coupling constant, $G_F$ [GeV$^{-2}$]. \\ Default: $0.116638 \times 10^{-4}$ GeV$^{-2}$. 
\end{description}

\subsubsection{Relevant Options for Process: \texttt{user}}
If process \texttt{user} is selected, i.e.\ a custom user-specified process, any
of the above physics options may be relevant if they are used in the
custom amplitude code. The code will have to be inspected to determine
this. The only built-in process-relevant option is:
\begin{description}[labelindent=1cm, labelwidth =\widthof{\bfseries9999999999999999999999}, leftmargin = !] 
	\item[\texttt{-M, -{}-mass <value>}] Relevant mass in the user specified process, $M$ [GeV]. Used in the scale choice and in the setup of kinematics. \\ Default: 0 GeV. 
\end{description}
Additional options may be added depending on the custom process/amplitude. See Section \ref{sec:newprocs} below for more details on the implementation of a custom process. 

\subsubsection{Loop Approximations} \label{sec:loopapprox} 
Small and large mass limits can be used as approximations for the quarks in the loop calculations. This requires some knowledge of the meaning of the approximations, hence this needs to be set by the user at compile time. 

In the file \texttt{input.f90} located in the \texttt{src}-directory, the subroutine \texttt{reset\_iloop\_array} can be found. This subroutine can be used by the user to set the \texttt{iloop\_array}, which is an array specifying the approximation used for each loop particle. The approximations that can be used are: 
\begin{description}[labelindent=1cm, labelwidth =\widthof{\bfseries9999999999999999999}, leftmargin = !] 
	\item[\texttt{iloop\_sm\_fermion}] Small mass limit for fermions. 
	\item[\texttt{iloop\_fm\_fermion}] Full mass effects for fermions. 
	\item[\texttt{iloop\_lm\_fermion}] Large mass limit for fermions. 
	\item[\texttt{iloop\_fm\_scalar}] Full mass effects for scalars. 
	\item[\texttt{iloop\_lm\_scalar}] Large mass limit for scalars. 
\end{description}
The size of the array must match the number of particles appearing in the loops, which should be checked by the user. Below is an example Fortran code snippet for the \texttt{reset\_iloop\_array} subroutine, which sets the loop approximations for an effective theory with SM top and bottom quarks, and one infinitely heavy top-partner: 
\begin{lstlisting}[language=Fortran, 
                   keywordstyle=\color{Red},
                   stringstyle=\color{Green},
                   identifierstyle=\color{Blue},
                   showstringspaces=false]
    if (size(iloop_array) == 3) then
      iloop_array = (/ iloop_fm_fermion, iloop_fm_fermion, &
                     & iloop_lm_fermion /)
    else
      call wae_error('reset_iloop_array', 'Expected size of &
           & iloop_array is 3, whereas actual one is', &
           & intval = size(iloop_array))
    end if
\end{lstlisting}

\subsubsection{Output} 

\begin{figure}[tbh] 
\centering
\begin{minipage}{.485\textwidth}
  \centering
  \includegraphics[width=\linewidth]{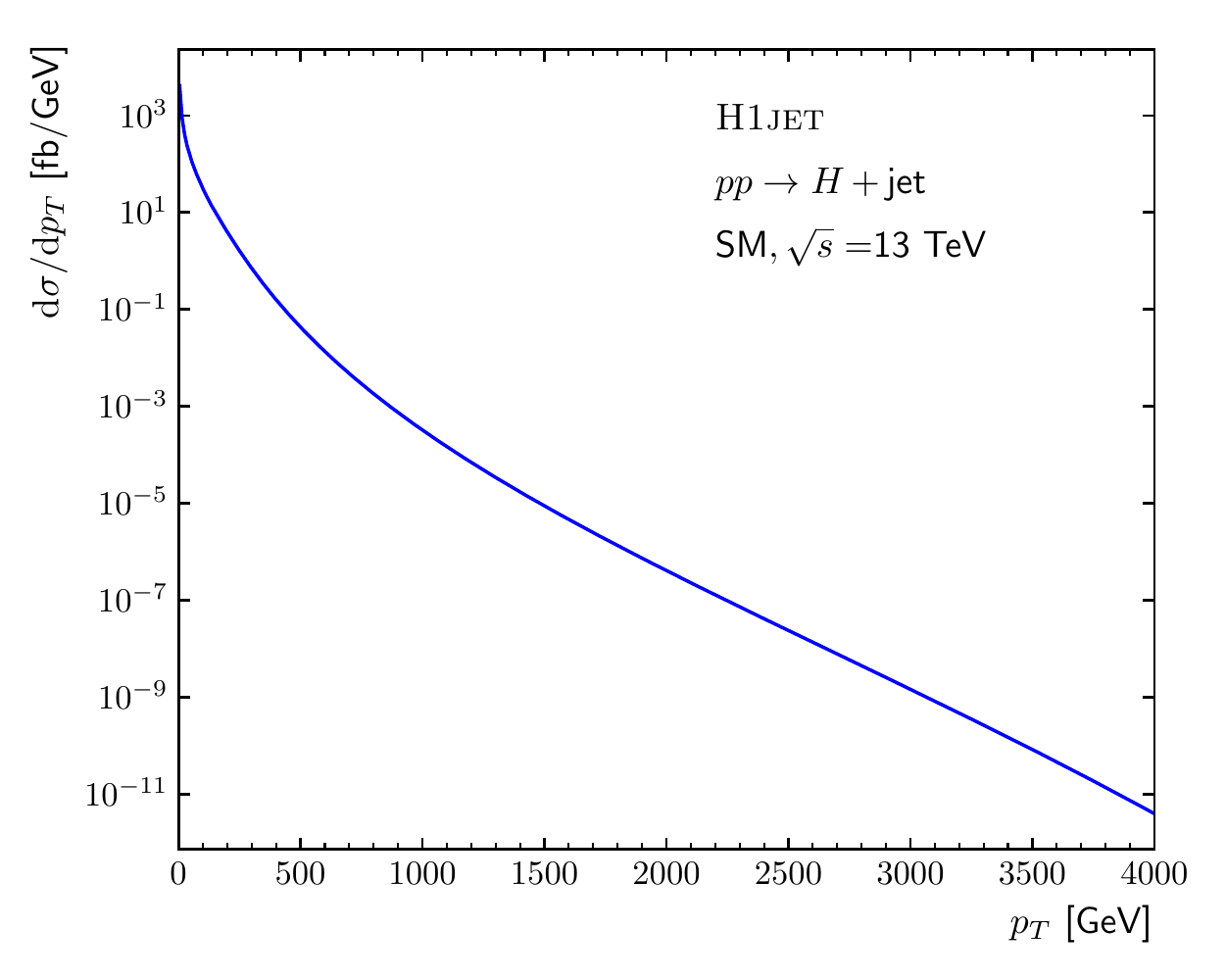}
  \captionof{figure}{The $p_{T}$ distribution for the SM process $pp \rightarrow H + \text{jet}$ from \textsc{H1jet} with default settings.}
  \label{fig:h1jetresult}
\end{minipage}%
\hfill%
\begin{minipage}{.485\textwidth}
  \centering
  \includegraphics[width=\linewidth]{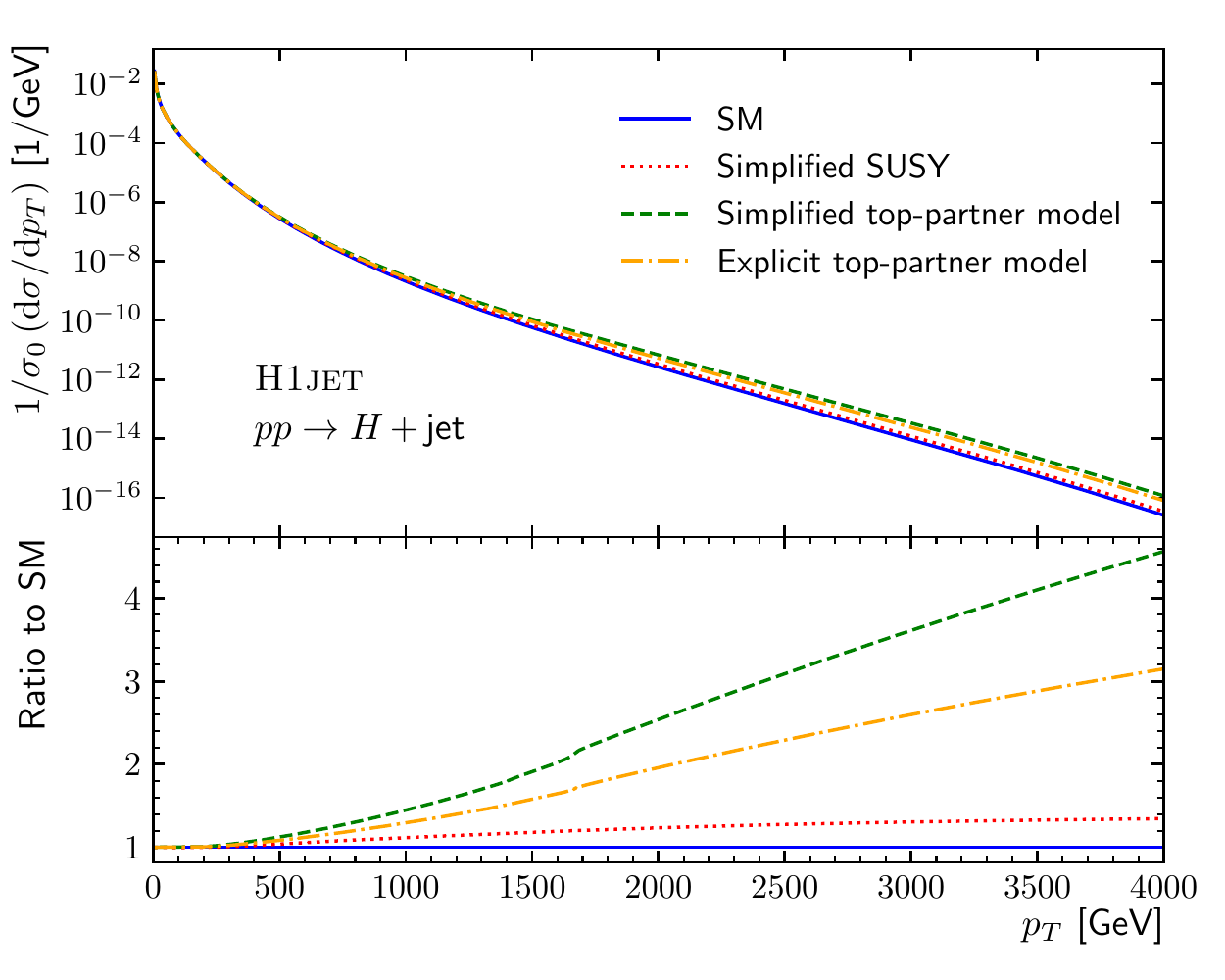}
  \captionof{figure}{Comparison between the CP-even Higgs $p_{T}$ distributions for various built-in models.}
  \label{fig:moneyplot}
\end{minipage}
\end{figure}

The helper script \texttt{PlotH1jet.py} facilitates easy and quick
plotting of the output from \textsc{H1jet}. The script requires Python
3 installed in order to run. The user needs to simply pipe the output
of \textsc{H1jet} to the script:
\begin{lstlisting}
	./bin/h1jet [options] | python PlotH1jet.py 
\end{lstlisting}
Alternatively, the plotting script can run on an output file from \textsc{H1jet}: 
\begin{lstlisting}
	./bin/h1jet [options] -o result.out 
	python PlotH1jet.py result.out 
\end{lstlisting}
A resulting example plot with default settings in \textsc{H1jet} is shown in \autoref{fig:h1jetresult}. 

A comparison between the various built-in models is shown in \autoref{fig:moneyplot}. Default SM parameters has been used with $m_{\tilde{t}_1} = 600$ GeV, $\Delta m = 200$ GeV, $\tan\beta = 5$, $m_T = 1.7$ TeV, $\sin^2\theta_{\tilde{t}} = \sin^2\theta_{T} = 0.1$, and $f = 900$ GeV, and considering the $\textbf{M4}_{\textbf{5}}$ model as the explicit top-partner model. 


\vspace{-3mm} 

\section{Benchmarking}
\label{sec:benchmarking}
\begin{figure}[tbh] 
\centering
\begin{minipage}{.485\textwidth}
  \centering
  \includegraphics[width=\linewidth]{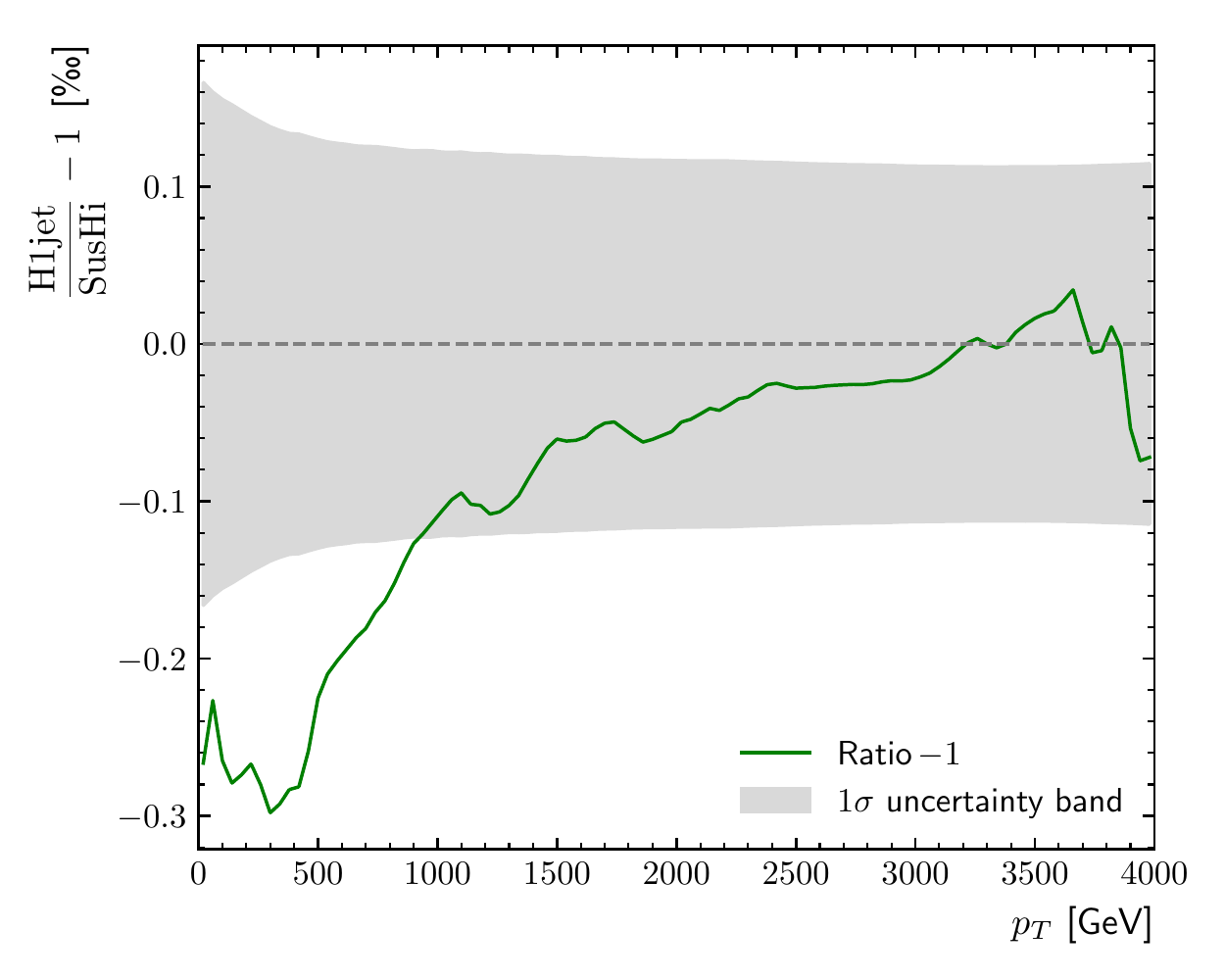}
  \captionof{figure}{The relative difference in permille between the \textsc{H1jet} result and that of \textsc{SusHi} for the $p_T$ distribution for a CP-odd Higgs. The grey band indicates the Monte Carlo error of \textsc{SusHi}.}
  \label{fig:cpoddratio}
\end{minipage}%
\hfill%
\begin{minipage}{.485\textwidth}
  \centering
  \includegraphics[width=\linewidth]{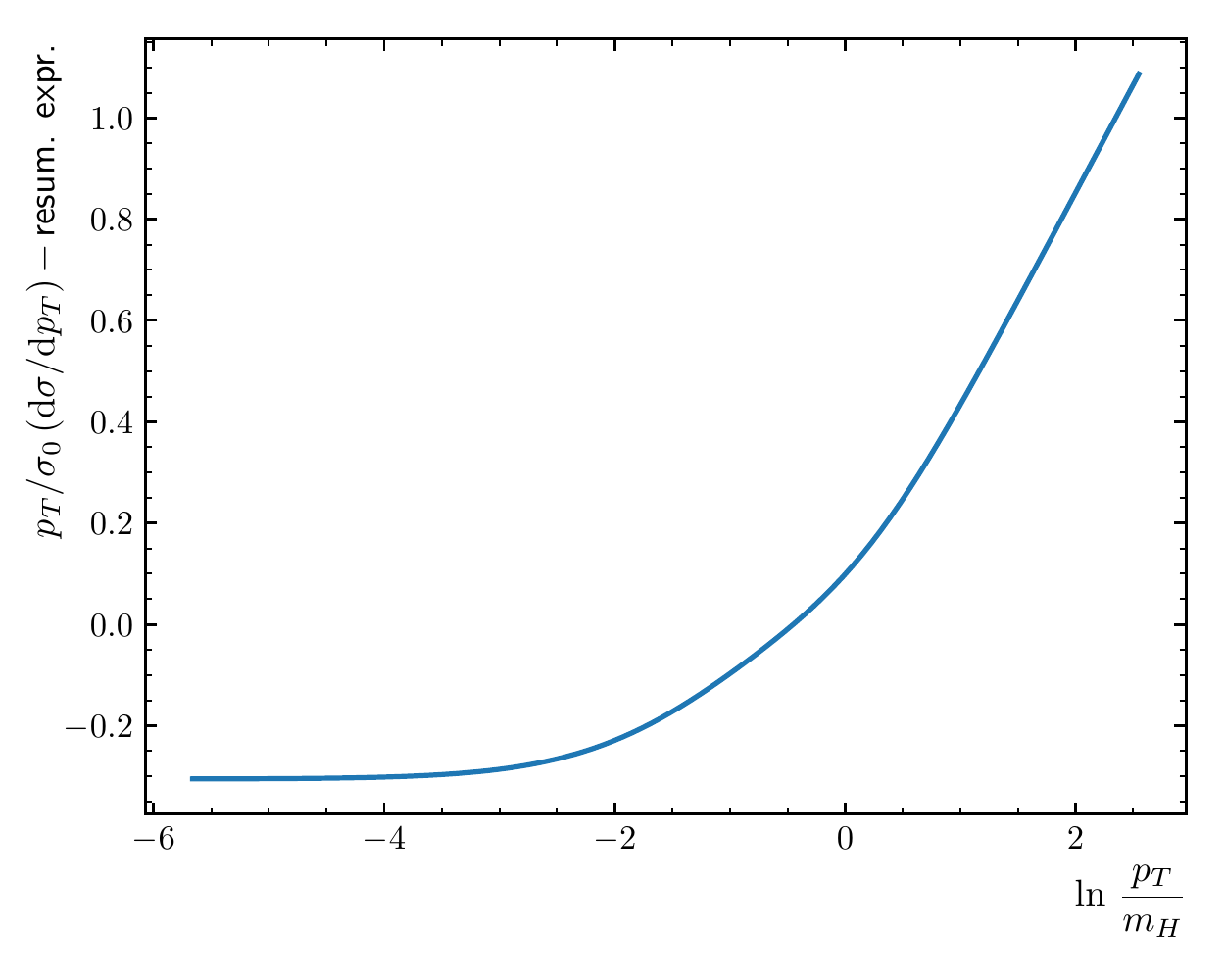}
  \captionof{figure}{The CP-odd distribution $\frac{p_T}{\sigma_0} \left ( \frac{\mathrm{d} \sigma}{\mathrm{d} p_T} \right )$ with the first term of the $p_T$ resummation subtracted, as a function of $\ln \frac{p_T}{m_H}$. For low $p_T$ values it convergences to a constant value.}
  \label{fig:cpoddresum}
\end{minipage}
\end{figure}

The various processes implemented in \textsc{H1jet} have been compared to those of \textsc{SusHi}~\cite{Harlander:2012pb,Harlander:2016hcx}, and have all been found to be in agreement. The relative ratio between the \textsc{H1jet} result and the
\textsc{SusHi} result for the $p_T$ distribution for the CP-odd Higgs
are shown in \autoref{fig:cpoddratio}, and is found to be in agreement
within the Monte Carlo error of \textsc{SusHi} for a large range of
$p_T$ values. Overall the agreement with \textsc{SusHi} is within
$3 \times 10^{-4}$. Note that the largest discrepancies were observed in the low $p_T$
region. To validate the \textsc{H1jet} results we have compared them
to the approximate expression valid at low $p_T$
\begin{equation}
	\frac{d \sigma}{d p_T} \xrightarrow{p_T \, \rightarrow \, 0} \sigma_0 \left [ 4 \, C_A \frac{\alpha_s}{\pi} \frac{1}{p_T} \left(\ln \frac{m_H}{p_T} + \mathcal{O} (1)\right) \right ] \,, \label{eq:resum}
\end{equation}
where $\sigma_0$ is the total Born cross section for $gg \rightarrow H$. In \autoref{fig:cpoddresum}, we show $\frac{p_T}{\sigma_0} \left ( \frac{\mathrm{d} \sigma}{\mathrm{d} p_T} \right )$ with the first term of eq.~\eqref{eq:resum} subtracted, as a function of $\ln \frac{p_T}{m_H}$. For $p_T \rightarrow 0$ this goes nicely towards a constant as expected. 

\begin{figure}[tbh] 
\centering
\begin{minipage}{.485\textwidth}
  \centering
  \includegraphics[width=\linewidth]{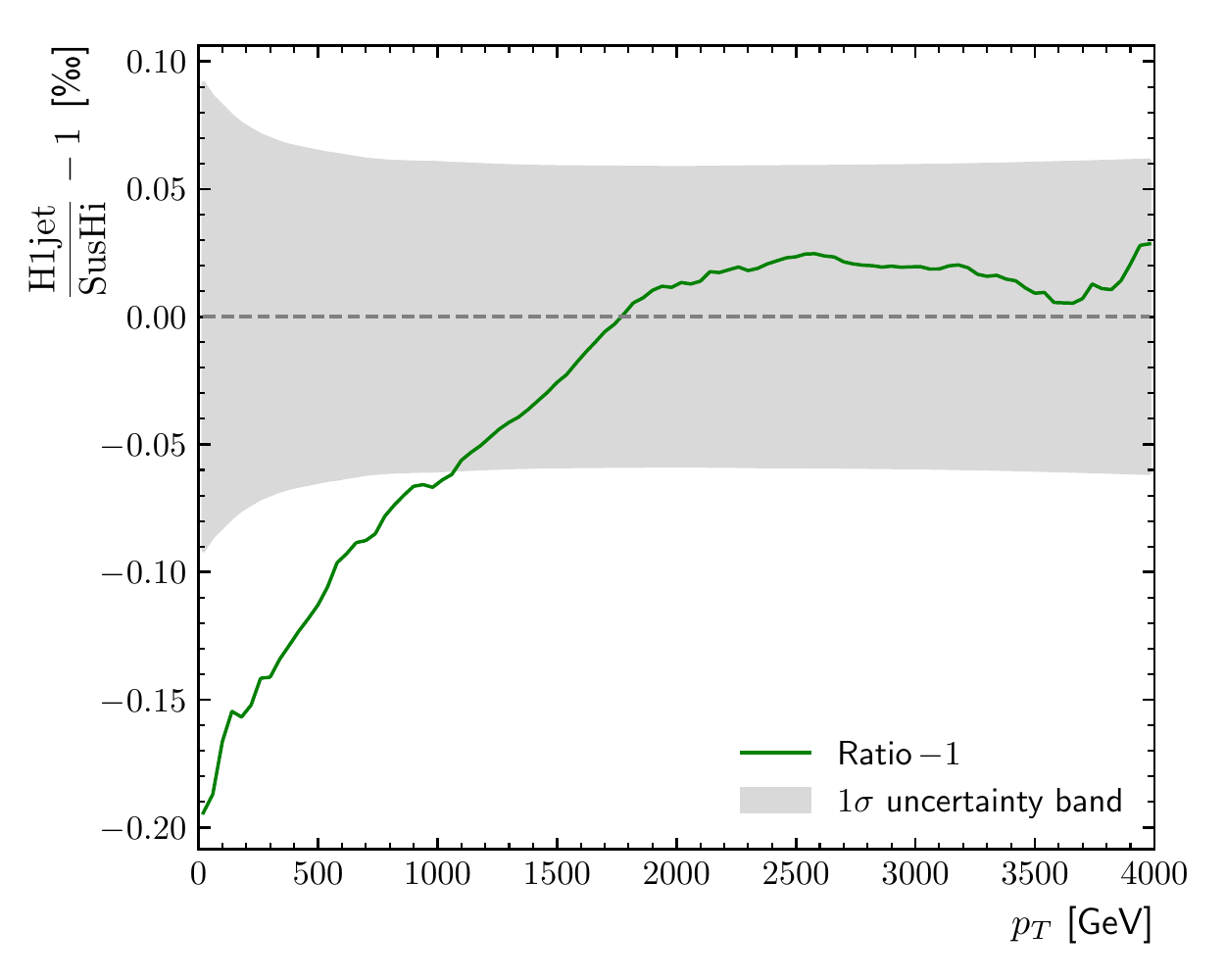}
  \captionof{figure}{The relative difference in permille between the \textsc{H1jet} result and that of \textsc{SusHi} for the $p_T$ distribution in SUSY. The grey band indicates the Monte Carlo error of \textsc{SusHi}.}
  \label{fig:susyratio}
\end{minipage}%
\hfill%
\begin{minipage}{.485\textwidth}
  \centering
  \includegraphics[width=\linewidth]{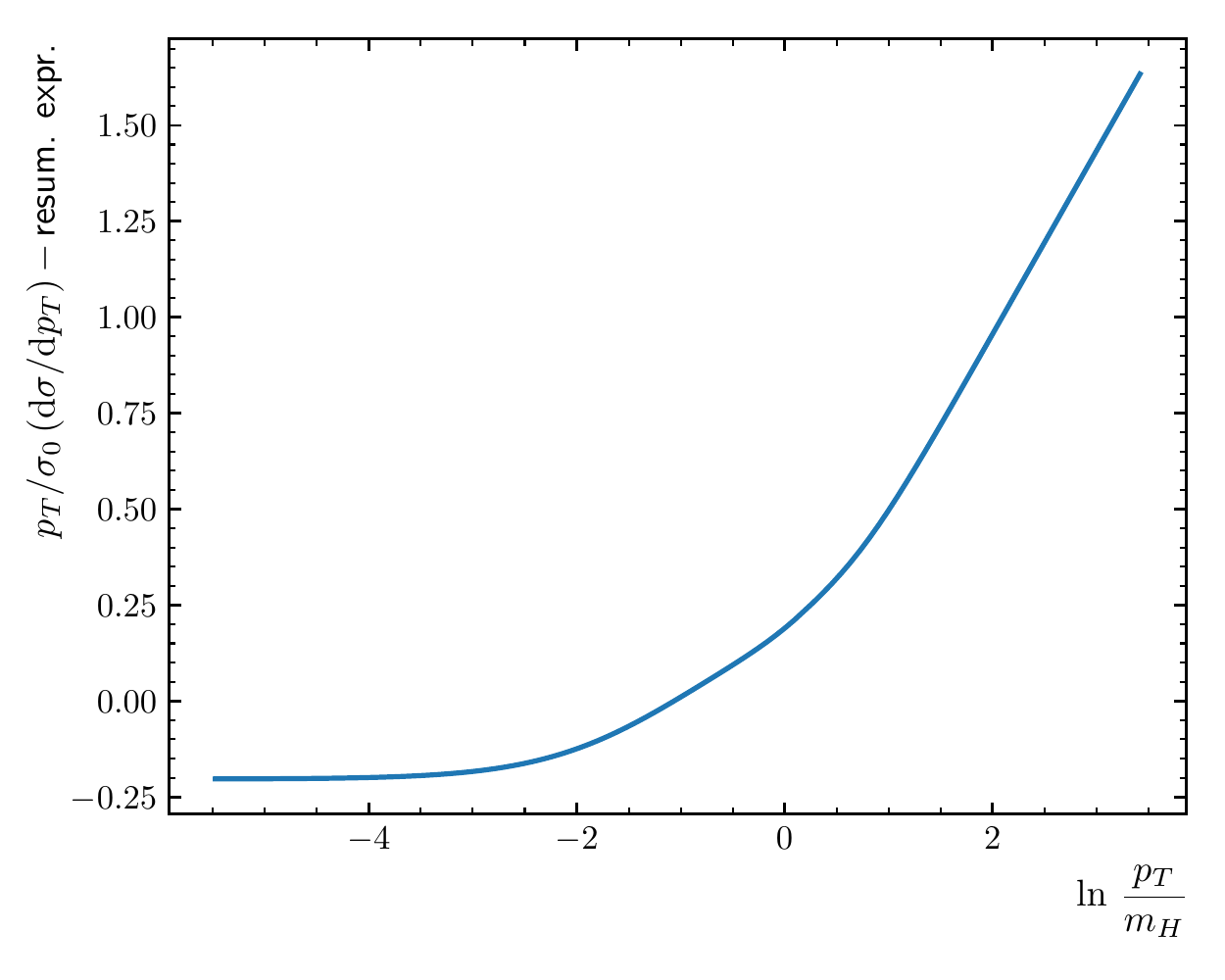}
  \captionof{figure}{The SUSY distribution $\frac{p_T}{\sigma_0} \left ( \frac{\mathrm{d} \sigma}{\mathrm{d} p_T} \right )$ with the first term of the $p_T$ resummation subtracted, as a function of $\ln \frac{p_T}{m_H}$. For low $p_T$ values it convergences to a constant value.}
  \label{fig:susyresum}
\end{minipage}
\end{figure}

The relative ratio between the \textsc{H1jet} and \textsc{SusHi} results for the SUSY are shown in \autoref{fig:susyratio} and is within $2 \times 10^{-4}$. Again the low $p_T$ behaviour can be checked by comparing to the resummed expression in~\autoref{fig:susyresum}. 

Note that our numerical accuracy crucially depends not only on the
accuracy of the numerical integration, but of that of the auxiliary
programs used to compute the PDF evolution (\textsc{HOPPET}) and the
scalar integrals (\textsc{CHAPLIN}). We have modified various internal
parameters of the two libraries, and we obtained differences that are
less than permille level. So, a conservative estimate of the numerical
uncertainty of \textsc{H1jet} is $1\times 10^{-3}$.

\section{Adding New Processes to \textsc{H1jet}}
\label{sec:newprocs} 
\textsc{H1jet} can be interfaced to use the squared matrix element
evaluated from a custom Fortran code. The implementation may be most
easily explained with a specific example. This section should be read very 
carefully before attempting to use the interface.

\subsection{Example: Axion-Like-Particle (ALP) Effective Theory}
We will present here a specific example of adding to \textsc{H1jet}
the production of a light axion-like-particle (ALP), $a$, along with a
jet. For simplicity, we only consider the gluon-fusion channel,
\begin{equation}
    g g \rightarrow g a \,. 
\end{equation}
This is a tree-level process due to an effective ALP-gluon coupling, 
\begin{equation}
   \delta \mathcal{L}_a \supset -c_{\tilde{G}}\frac{a}{f_a} G_{\mu\nu}^{a} \tilde{G}^{a \mu\nu} \,, \label{eq:ggacoupling}
\end{equation}
where $G_{\mu\nu}^{a}$ is the gluon field strength tensor and
$\tilde{G}^{a}_{ \mu\nu}=(1/2)\epsilon_{\mu\nu\rho\sigma}{G}^{a
  \rho\sigma}$ its dual.  The model and the
\textsc{FeynRules}~\cite{Alloul:2013bka} model files are described and
provided in ref.~\cite{Brivio:2017ije}. We use
\textsc{FeynCalc}~\cite{Mertig:1990an,Shtabovenko:2016sxi,Shtabovenko:2020gxv}
to evaluate the amplitude from the model, so we have to convert the
\textsc{FeynRules} model to a \textsc{FeynArts}~\cite{Hahn:2000kx}
model in Mathematica:
\begin{mmaCell}{Code}
<< FeynRules` 
\end{mmaCell}
\begin{mmaCell}{Code}
LoadModel["SM.fr", "alp_linear.fr", "alp_linear_operators.fr"];  
\end{mmaCell}
\begin{mmaCell}{Code}
WriteFeynArtsOutput[LSM + LALP, CouplingRename -> False]; 
\end{mmaCell}
The resulting \textsc{FeynArts} model files are written to a new directory \texttt{ALP\_linear\_FA}, which needs to be moved to the \textsc{FeynArts/Models} directory. Note that in the \textsc{FeynArts} model, the ALP field is called \texttt{S[4]} and the gluon fields are called \texttt{V[4]}. 

In a new Mathematica session, we load \textsc{FeynCalc} with \textsc{FeynArts}: 
\begin{mmaCell}{Code}
$LoadAddOns = {"FeynArts"}; 
\end{mmaCell}
\begin{mmaCell}{Code}
<< FeynCalc` 
\end{mmaCell}
The \texttt{FeynArts/Models} directory can be located with: 
\begin{mmaCell}{Code}
$FeynArtsDir
\end{mmaCell}
First, we patch the \texttt{ALP\_linear\_FA} model with: 
\begin{mmaCell}{Code}
FAPatch[PatchModelsOnly -> True] 
\end{mmaCell}
This ensures that the model files works with \textsc{FeynCalc}. 

Then we create the tree-level $2 \rightarrow 2$ topologies and insert
the relevant fields for our process:
\begin{mmaCell}{Code}
tops = CreateTopologies[0, 2 -> 2];
\end{mmaCell}
\begin{mmaCell}{Code}
ins = InsertFields[tops, {V[4], V[4]} -> {V[4], S[4]}, InsertionLevel -> {Classes}, Model -> "ALP_linear_FA", GenericModel -> "ALP_linear_FA"];
\end{mmaCell}
It is possible to draw the Feynman diagrams for the process as a check: 
\begin{mmaCell}{Code}
Paint[ins, ColumnsXRows -> {2, 1}, Numbering -> Simple, SheetHeader -> None, 
      ImageSize -> {512, 256}]; 
\end{mmaCell}
We then set up the amplitude: 
\begin{mmaCell}{Code}
feynamp = CreateFeynAmp[ins];  
\end{mmaCell}
\begin{mmaCell}{Code}
amp = FCFAConvert[feynamp, IncomingMomenta -> {k1, k2}, OutgoingMomenta -> {k3, k4}, 
                  UndoChiralSplittings -> True, ChangeDimension -> 4, 
                  TransversePolarizationVectors -> {k1, k2, k3}, List -> False, 
                  SMP -> True, Contract -> True, DropSumOver -> True] 
\end{mmaCell}
While not strictly necessary, it is recommended to enable the \texttt{SMP} option. Any additional substitutions in the amplitude can be specified with the \texttt{FinalSubstitutions} option. 

We then set up the kinematics: 
\begin{mmaCell}{Code}
FCClearScalarProducts[]; 
\end{mmaCell}
\begin{mmaCell}{Code}
SetMandelstam[s, t, u, k1, k2, -k3, -k4, 0, 0, 0, mA]; 
\end{mmaCell}
We introduce here a parameter \texttt{mA} for the ALP mass $m_a$. 

We then square the amplitude: 
\begin{mmaCell}{Code}
ampsquared = Simplify[
  (TrickMandelstam[#1, {s, t, u, mA^2}] & )[
    (DoPolarizationSums[#1, k2, k1, ExtraFactor -> 1/2] & )[
      (DoPolarizationSums[#1, k1, k2, ExtraFactor -> 1/2] & )[ 
        (DoPolarizationSums[#1, k3, 0] & )[ 
          (SUNSimplify[#1, Explicit -> True, SUNNToCACF -> False] & )[
            FeynAmpDenominatorExplicit[(1 / (SUNN^2 - 1)^2) * (amp * ComplexConjugate[amp])]]]]]]] /. SUNN -> 3 
\end{mmaCell}
Setting the \texttt{SUNNToCACF} option in \texttt{SUNSimplify[]} to \texttt{False} is not necessary, nor is it necessary to fix \texttt{SUNN} to $3$. This can be handled by the dressing script and \textsc{H1jet}. 

Finally, we  write the amplitude as Fortran code to a file: 
\begin{mmaCell}{Code}
Write2["ALP_amp.f90", gg = ampsquared, FormatType -> FortranForm, FortranFormatDoublePrecision -> False] 
\end{mmaCell}
Note here that we specify the gluon-gluon channel with the \texttt{gg = ampsquared} input to the function. This is required for the subsequent dressing script to work properly. It is important to specify the $2$-particle initial state by using combinations of \texttt{g}, \texttt{u}, \texttt{d}, \texttt{c}, \texttt{s}, \texttt{b}, \texttt{ubar}, \texttt{dbar}, \texttt{cbar}, \texttt{sbar}, and \texttt{bbar}. One can also use \texttt{q} and \texttt{qbar} for all the light quarks and antiquarks respectively, i.e.\ $u$, $d$, $c$, and $s$. For example, \texttt{bbbar} will be the $b\bar{b}$ channel. \\ 

The generated Fortran code \texttt{ALP\_amp.f90} has the following content:
\begin{lstlisting}[language=Fortran,keywordstyle=\color{Red},stringstyle=\color{Green},identifierstyle=\color{Blue},showstringspaces=false]
        gg = (3.*CGtil**2*(mA**8 - 2.*mA**6*(t + u) + 
     &      3.*mA**4*(t + u)**2 - 2.*mA**2*(t + u)**3 + 
     &      (t**2 + t*u + u**2)**2)*SMP("g_s")**2)/
     &  (fa**2*s*t*u)
\end{lstlisting}
This code has to be dressed by the Python helper script \texttt{DressUserAmpCode.py}: 
\begin{lstlisting}
	python DressUserAmpCode.py ALP_amp.f90 
\end{lstlisting}
This produces a dressed Fortran code file called by default
\texttt{user\_interface.f90}.

The helper script provides a help message which can be called with
\texttt{-h} or \texttt{-{}-help}. The name of the output file can be
specified with the \texttt{-o} option. Multiple input Fortran files
can be given as arguments to the helper script. The full usage is:
\begin{lstlisting}
	python DressFeynCalcCode.py [-h] [-o [OUTFILE]] inputfile [inputfile ...] 
\end{lstlisting}
The provided input Fortran code files does not necessarily have to be
generated with \textsc{FeynCalc}. They can be generated by any other
program or even be written by hand by providing the appropriate
expression for \texttt{gg} or the desired channel.

To use the new dressed custom Fortran code with \textsc{H1jet}, it is necessary to recompile \textsc{H1jet} with the custom Fortran code: 
\begin{lstlisting}
	./configure USERFILE=/path/to/custom/user_interface.f90 
	make clean
	make 
\end{lstlisting}
Running \texttt{./h1jet -{}-help} we see that three new additional options have been added: 
\begin{description}[labelindent=1cm, labelwidth =\widthof{\bfseries9999999999999999999999}, leftmargin = !] 
	\item[\texttt{-{}-c\_CGtil <value>}] The Wilson coefficient $c_{\tilde{G}}$ in eq.~\eqref{eq:ggacoupling}. 
	\item[\texttt{-{}-c\_mA <value>}] The ALP mass, $m_a$. 
	\item[\texttt{-{}-c\_fa <value>}] The ALP suppression scale $f_a$ in eq.~\eqref{eq:ggacoupling}. 
\end{description}
The leading \texttt{c\_} in the name stands for ``custom'' and is automatically added in order to avoid naming issues in the code. \\ 

\begin{figure}[tbh] 
\centering
\begin{minipage}{.485\textwidth}
  \centering
  \includegraphics[width=\linewidth]{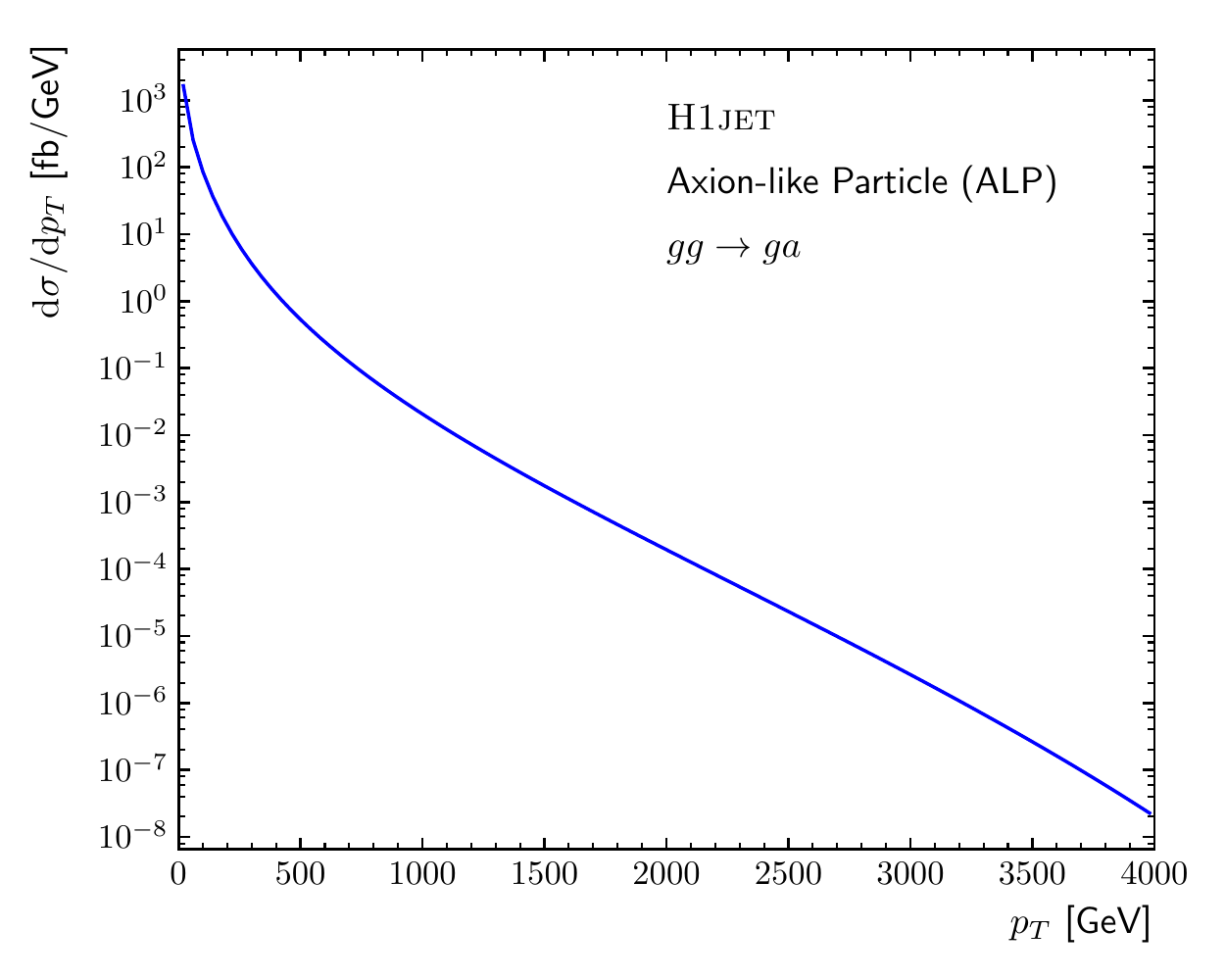}
  \captionof{figure}{The $p_{T}$ distribution for the process $gg \rightarrow ga$ from \textsc{H1jet} with the amplitude from the user interface.}
  \label{fig:alpresult}
\end{minipage}%
\hfill%
\begin{minipage}{.485\textwidth}
  \centering
  \includegraphics[width=\linewidth]{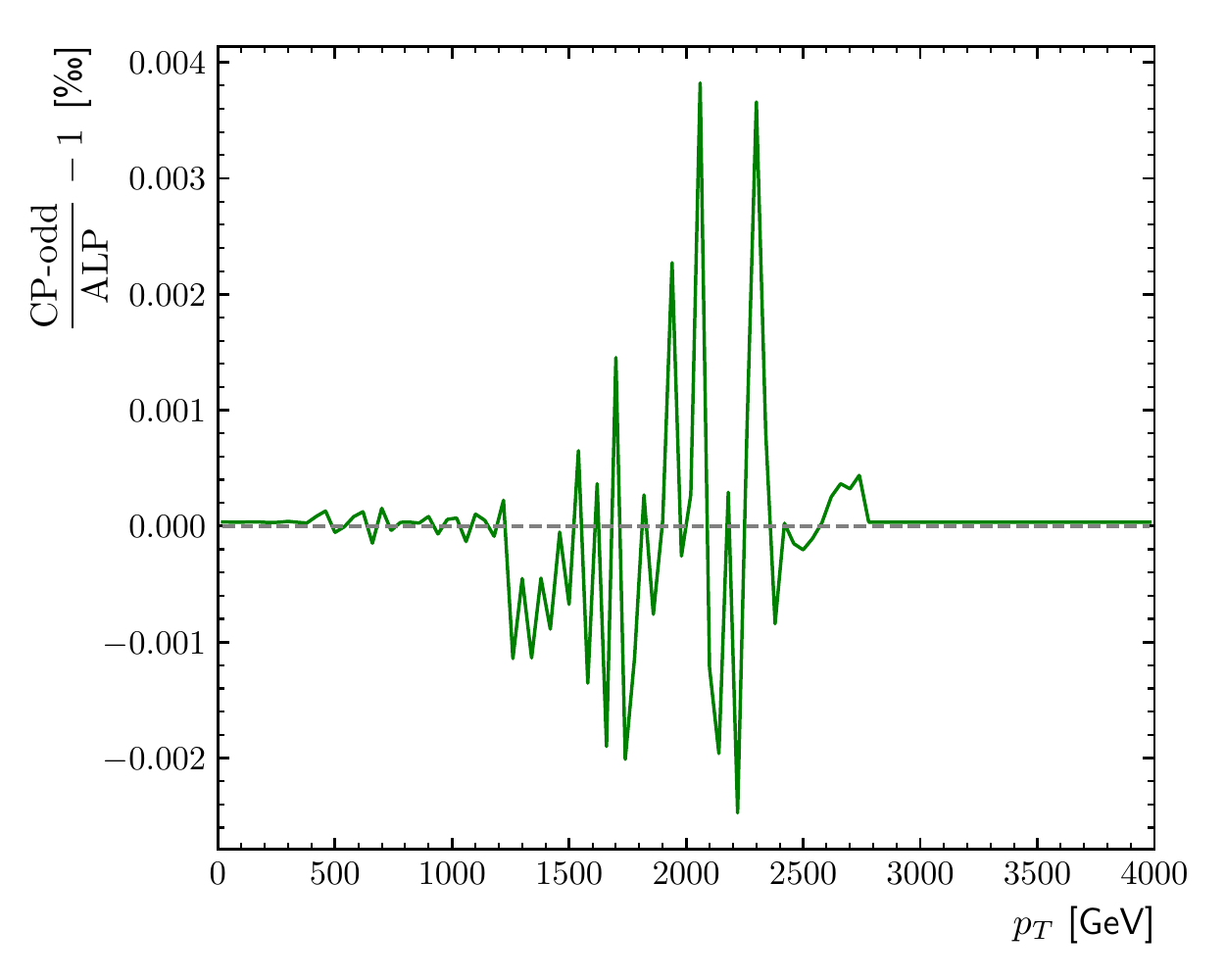}
  \captionof{figure}{The relative ratio in permille between the CP-odd result and the result with the amplitude from the user interface.}
  \label{fig:alpratio}
\end{minipage}
\end{figure}

The result from the ALP implementation in \textsc{H1jet} is shown in \autoref{fig:alpresult} and can be compared to the \textsc{H1jet} result for the CP-odd Higgs by using a single top quark in the loop with an infinite mass limit, resulting in an effective coupling between the CP-odd Higgs and the gluons. In fact, the respective ALP and CP-odd couplings are then related as such, 
\begin{equation}
	\frac{c_{\tilde{G}}^2}{f_a^2} = \frac{\alpha_s^2}{64 \pi^2 v^2} \,. 
\end{equation}
The comparison is shown in \autoref{fig:alpratio}, where we see agreement within $4\times 10^{-6}$. 

The result can also be compared to the same \textsc{FeynRules} model
used with \textsc{\linebreak MadGraph5\_aMC@NLO}~\cite{Alwall:2014hca}, where our
code takes one second to run, while \textsc{MadGraph} can take up to
several hours depending on the number of events, due to
\textsc{MadGraph} running a full Monte Carlo integration. We have
found that \textsc{H1jet} agrees with \textsc{MadGraph5\_aMC@NLO}
within Monte Carlo errors. We have also seen that
\textsc{MadGraph5\_aMC@NLO} runs into numerical instabilities at low
$p_T$, while \textsc{H1jet} has by construction the correct behaviour.

\subsubsection{The Total Cross Section} \label{sec:alptotxsec} 
While not strictly necessary for the user interface to run, it is still recommended to add the code for the evaluation of the total cross section to the custom user interface. This is easy to do as well. We will here show it for the ALP model. 

We start with considering the gluon-fusion ALP production, $gg \rightarrow a$. In Mathematica, create a tree-level $2 \rightarrow 1$ topology, and insert the fields: 
\begin{mmaCell}{Code}
tops = CreateTopologies[0, 2 -> 1];
\end{mmaCell}
\begin{mmaCell}{Code}
ins = InsertFields[tops, {V[4], V[4]} -> {S[4]}, InsertionLevel -> {Classes}, Model -> "ALP_linear_FA", GenericModel -> "ALP_linear_FA"];
\end{mmaCell}
Then we set up the amplitude: 
\begin{mmaCell}{Code}
feynamp = CreateFeynAmp[ins]; 
\end{mmaCell}
\begin{mmaCell}{Code}
amp = FCFAConvert[feynamp, IncomingMomenta -> {k1, k2}, 
                  OutgoingMomenta -> {k3}, UndoChiralSplittings -> True, 
                  ChangeDimension -> 4, TransversePolarizationVectors -> {k1, k2}, 
                  List -> False, SMP -> True, Contract -> True, DropSumOver -> True] 
\end{mmaCell}
as well as the kinematics: 
\begin{mmaCell}{Code}
FCClearScalarProducts[]; 
\end{mmaCell}
\begin{mmaCell}{Code}
SP[k1, k1] = 0; 
\end{mmaCell}
\begin{mmaCell}{Code}
SP[k2, k2] = 0; 
\end{mmaCell}
\begin{mmaCell}{Code}
SP[k3, k3] = mA^2; 
\end{mmaCell}
\begin{mmaCell}{Code}
SP[k1, k2] = mA^2 / 2; 
\end{mmaCell}
We then square the amplitude: 
\begin{mmaCell}{Code}
ampsquared = Simplify[
  (DoPolarizationSums[#1, k2, k1, ExtraFactor -> 1/2] & )[
    (DoPolarizationSums[#1, k1, k2, ExtraFactor -> 1/2] & )[ 
      (SUNSimplify[#1, Explicit -> True, SUNNToCACF -> False] & )[
        FeynAmpDenominatorExplicit[(1 / (SUNN^2 - 1)^2) * (amp * ComplexConjugate[amp])]]]]]
\end{mmaCell}
For a $2 \rightarrow 1$ process, the hadronic cross section is given in eq.~\eqref{eq:Born-had}, where the partonic luminosity $\mathcal{L}_{gg}$ is handled by \textsc{H1jet}. Hence, we need to multiply our squared matrix element by
\begin{mmaCell}{Code}
xsec = Pi * ampsquared / mA^4 
\end{mmaCell}
Finally, we can write the cross section as a Fortran code file: 
\begin{mmaCell}{Code}
Write2["ALP_xsec.f90", xsgg = xsec, FormatType -> FortranForm, FortranFormatDoublePrecision -> False] 
\end{mmaCell}
We again specify the gluon-gluon channel with the \texttt{xsgg = xsec}, but this time indicate with the leading \texttt{xs} that the code is for the Born cross section. Otherwise, the same rules apply. It is crucial to make sure not to save the Born cross section in the same file as the squared amplitude code for the transverse momentum distribution. 

The new generated code \texttt{ALP\_xsec.f90} is provided to \texttt{DressUserAmpCode.py} along with the squared amplitude code: 
\begin{lstlisting}
	python DressUserAmpCode.py ALP_amp.f90 ALP_xsec.f90 
\end{lstlisting}
And \textsc{H1jet} can be recompiled to include the new source code: 
\begin{lstlisting}
	./configure USERFILE=/path/to/custom/user_interface.f90 
	make clean
	make 
\end{lstlisting}
After which \textsc{H1jet} will calculate the Born cross section for the custom process. 

\section{Conclusions}
\label{sec:the-end}

We have presented a method that allows a fast computation of the
transverse momentum distribution of a colour singlet. The method is
implemented in the program \textsc{H1jet}, which returns a transverse momentum
spectrum for the specified colour singlet in about a second. \textsc{H1jet} is
similar in spirit to \textsc{SusHi}, but is incomparably faster.

The program implements various processes, including Higgs production
both in gluon fusion and bottom-antibottom annihilation, as well as
$Z$ production. Loop-induced Higgs production is implemented not only
in the SM, but also in attractive BSM scenarios, such as SUSY or
composite Higgs. For SUSY, we implement a simplified model with two
stops, as done in ref.~\cite{Banfi:2018pki}. For composite Higgs, we
implement both the simplified model of ref.~\cite{Banfi:2013yoa}, as
well as some explicit models with one or more top
partners~\cite{Banfi:2019xai}. Loop integrals can be computed either
exactly or in the infinite-mass limit. The latter limit implements in
practice a dimension-6 contact interaction between the Higgs and the
gluon field. The program is very flexible, and the only
process-dependent input is the corresponding amplitude in terms of
Mandelstam invariants. This can be computed by the user either
manually, or with the use of automated programs such as
FeynCalc~\cite{Shtabovenko:2020gxv}, and connected to the program via
a simple interface. As an example, we have included in the package the
calculation of the transverse momentum distribution of an ALP starting
from the general Feynman rules of ref.~\cite{Brivio:2017ije}. Note
that the possibility of automatically implementing a new model inside
the program is a feature that is not available in \textsc{SusHi}. We
also stress that it is also possible to take advantage of input files
to obtain results for an arbitrary number of fermions and scalars in
loops, with appropriate couplings. This could be used, for instance, to
implement the MSSM instead of the provided simplified SUSY model.

We stress that \textsc{H1jet} is not a replacement for a proper Monte Carlo
analysis implementing realistic experimental cuts. However, we believe
it will be invaluable for BSM experts to assess whether a given model
gives sizeable deviations from the SM. In fact, due to its fast
implementation, \textsc{H1jet} makes it possible to perform parameter scans in
seconds, and to take into account mass effects in specific
models. Also, due to the fact that \textsc{H1jet} is not based on a Monte Carlo
integration, one can separate interference between different
contributions very precisely, something which is very difficult to
achieve with Monte Carlo event generators.

\textsc{H1jet} can be also useful to precision phenomenology. In fact, it makes
it possible to easily perform theoretical studies of the transverse
momentum distribution of a colour singlet, especially those involving
the matching of resummed calculations with exact fixed order. Also,
although the implemented cross sections are computed at the lowest
order in QCD, nothing prevents the inclusion of higher orders, provided one
integrates over all coloured particles. 

\paragraph{Acknowledgements.} The idea of having a fast program to
compute transverse momentum spectra started from AB's collaboration
with G.~Zanderighi, P.~F.~Monni and F.~Caola. AB
acknowledges many useful discussions with them on the topic. AB also
wishes to thank B.~Dillon, W.~Ketaiam and S.~Kvedarait\.e for contributions to a
private preliminary version of \textsc{H1jet}. We thank J.~M.~Lindert for all his 
helpful remarks and suggestions on this paper. The studentship of AL is
supported by the Science Technology and Facilities Council (STFC)
under grant number ST/P000819/1. The work of AB is supported by the
Science Technology and Facilities Council (STFC) under grants number
ST/P000819/1 and ST/T00102X/1.

\appendix

\section{Implementation of Scalar Integrals}
\label{sec:scalar-integrals}

This appendix contains the details of how \textsc{H1jet} computes one-loop
scalar integrals that are relevant for Higgs production. These
integrals depend on one internal mass, which we denote by $m$, and are
functions of Mandelstam
invariants.

Scalar integrals can be written in terms of logarithms and dilogarithms of complex
arguments, which require appropriate analytic continuations. Instead of performing such manipulations ourselves, we have
decided to use the implementation of the library CHAPLIN, and recast all relevant transcendental
functions into harmonic polylogarithms $H(\vec a;z)$, with
$\vec a=(a_1,\dots,a_n)$. For real values of the argument of a polylogarithm, CHAPLIN
uses the $+i\varepsilon$ prescription, i.e.\ $H(\vec a;z)$ with $z$
real is interpreted as $H(\vec a;z+i\varepsilon)$. Therefore, we need
to make sure that the imaginary part of the argument of scalar
integrals is consistent with the convention of CHAPLIN.

The relevant one-loop integrals we need to deal with are  
bubbles, triangles and boxes.

\paragraph{Bubbles.} The bubble integral is defined as
\begin{equation}
  \label{eq:bubble}
  B_0(s)=2 -\sqrt{1-\frac{4(m^2-i\varepsilon)}{s}}\ln\left[-\frac{z}{1-z}\right]\,,
\end{equation}
where 
\begin{equation}
  \label{eq:z}
  z=\frac{1}{2}\left(1+\sqrt{1-\frac{4(m^2-i\varepsilon)}{s}}\right)\,.
\end{equation}
The argument of the logarithm in eq.~\eqref{eq:bubble} has a different
form according to the value of $s$:
\begin{equation}
  \label{eq:z-omz}
  -\frac{z}{1-z} = \left\{
    \begin{split}
      &-\frac{1+\sqrt{1+\frac{4m^2}{|s|}}}{1-\sqrt{1+\frac{4m^2}{|s|}}}\,,\qquad \qquad s<0\,,\\
        & -\frac{1+i\sqrt{\frac{4m^2}{s}-1}}{1-i\sqrt{\frac{4m^2}{s}-1}}\,,\qquad \quad\>\> 0<s<4m^2\,,\\
        &
        -\frac{1-\sqrt{1-\frac{4m^2}{s}}}{1-\sqrt{1-\frac{4m^2}{s}}}-i\varepsilon\,,\qquad
        s>4m^2
    \end{split}
\right.
\end{equation}
Note that the only case in which one needs a small imaginary part is
the case $s>4m^2$. This imaginary part has the opposite convention as
in CHAPLIN. As a solution, we invert the argument of the logarithm and
use the identity $\ln z = -\ln(1/z)$. In practice, after an appropriate analytic
continuation of the square root, we define
\begin{equation}
  \label{eq:z-implemented}
  z \equiv \left\{\begin{split}
      \frac{1}{2}\left(1+\sqrt{1-\frac{4m^2}{s}}\right)\,,& \qquad \frac{4m^2}{s}<1\\
      \frac{1}{2}\left(1+i\sqrt{\frac{4m^2}{s}-1}\right)\,,& \qquad \frac{4m^2}{s}>1
    \end{split}
    \right.
  \end{equation}
and implement the bubble as follows:
\begin{equation}
  \label{eq:bubble-implemented}
  B_0(s)=
      2 -(2 z-1) \,H\left(1;\frac{1}{z}\right) \,.
\end{equation}
Note that a logarithm of a negative number can also be correctly
analytically continued by using the default Fortran implementation
of the complex logarithm. As for CHAPLIN, Fortran assumes that a
negative number has a small positive imaginary part. Therefore, in
case we have a small negative imaginary part, we can still use the
relation $\ln z = -\ln(1/z)$, which gives the correct analytic continuation.

\paragraph{Triangles.} The triangle integral $C_0(s)$ is defined as
\begin{equation}
  \label{eq:triangle}
  C_0(s) = \frac{1}{2s}\ln^2\left[-\frac{z}{1-z}\right]\,,
\end{equation}
where $z$ is given in eq.~\eqref{eq:z}. Again, for $s>4m^2$, the argument of the logarithm has the opposite sign with respect to what is implicitly assumed by CHAPLIN. Therefore, we invert again the argument of the logarithm, and using the definition of $z$ in eq.~\eqref{eq:z-implemented}, we implement the triangle as follows:
\begin{equation}
  \label{eq:triangle-HPL}
  s\,C_0(s) = H\left(1,1;\frac{1}{z}\right)\,.
\end{equation}

\paragraph{Boxes.} The scalar four-point function with three massless
(the gluons) and one massive (the Higgs boson) external lines is given
by~\cite{Baur:1989cm},
\begin{multline}
  \label{eq:box}
	D(s, t) = \frac{1}{st} \int_{0}^{1} \frac{\mathrm{d}x}{x(1-x)+m^2u/(ts)} \left [- \ln\left ( 1 - i \varepsilon - \frac{m_H^2}{m^2} x (1-x) \right )\right. \\ \left.
	+ \ln\left ( 1 - i \varepsilon - \frac{s}{m^2} x (1-x) \right ) + \ln\left ( 1 - i \varepsilon - \frac{t}{m^2} x (1-x) \right ) \right ] \,, 
\end{multline}
which can be expressed in terms of complex dilogarithms by using the exact result
\begin{multline}
  \label{eq:ci-dilogs}
  \frac{1}{st} \int_{0}^{1} \frac{dx}{x(1-x)+m^2u/(ts)} \ln\left ( 1 - i \varepsilon - \frac{v}{m^2} x (1-x) \right )= \frac{2}{\sqrt{1+4m^2u/(ts)}} \times  \\ \times
  \left[
    \mathrm{Li}_2\left(\frac{x_-}{x_--y}\right)-\mathrm{Li}_2\left(\frac{x_+}{x_+-y}\right)
    +\mathrm{Li}_2\left(\frac{x_-}{y-x_+}\right)-\mathrm{Li}_2\left(\frac{x_+}{y-x_-}\right)
    \right.
    \\ \left.
    + \ln\left( \frac{-x_{-}}{x_{+}} \right) \ln\left ( 1 - i \varepsilon - \frac{v}{m^2}x_-x_+ \right )
    \right]\,,
  \end{multline}
  where
  \begin{equation}
    \label{eq:xpm}
    	x_{\pm} \equiv \frac{1}{2} \left ( 1 \pm \sqrt{1 + \frac{4m^2 u}{ts}} \right )
  \end{equation}
are real numbers, with $x_{+} > 1$ and $x_{-} < 0$, and
\begin{equation}
  \label{eq:y-def}
  y \equiv \frac{1}{2} \left ( 1 + \sqrt{1 - \frac{4(m^2 - i \varepsilon)}{v}} \right ) \,,
\end{equation}
acquires an imaginary part according to the value of $v$. In particular, keeping track of the imaginary part of $y$ yields
\begin{equation}
  \label{eq:y-implementation}
y = \left\{
  \begin{split}
  \frac{1}{2}\left(1+\sqrt{1+\frac{4m^2}{|v|}}\right)-i\varepsilon\,,& \qquad v<0\\
  \frac{1}{2}\left(1+i\sqrt{\frac{4m^2}{v}-1}\right)\,,& \qquad 0<v<4 m^2\\
    \frac{1}{2}\left(1+\sqrt{1-\frac{4m^2}{v}}\right)+i\varepsilon\,,& \qquad v>4m^2
  \end{split}
  \right.
\end{equation}
From the above, we see that w for $0<v<4 m^2$ we can use the
dilogarithms as given by CHAPLIN.
For $v<0$, $x_-/(x_--y)$ and $x_+/(y-x_-)$ acquire a small positive
imaginary part, whereas $x_+/(x_+-y)$ and $x_-/(y-x_+)$ a small
negative imaginary part. The reverse happens for $v>4m^2$. Therefore,
we need to perform some formal manipulations to use the harmonic
polylogarithms provided by CHAPLIN.

In practice, whenever the argument $z$ of the dilogarithm is complex,
we just use the definitory relation $\mathrm{Li_2}(z)=H(0,1;z)$. When
$z=x+i\varepsilon$, with $x$ real, we use
$\mathrm{Li_2}(x+i\varepsilon)=H(0,1;x)$, with $H(0,1;x)$ the complex
number provided by CHAPLIN. If $z=x-i\varepsilon$, we use the
identities
\begin{equation}
  \begin{split}
\mathrm{Li}_{2}\left ( x-i\varepsilon \right )&= -H\left(0,1;\frac{1}{x} \right ) - \frac{\pi^2}{6} - H(0,0; -x)  = -H(0,-1;-x) = \\ & 
= -H\left (0,1, 1-x \right )+ \frac{\pi^2}{6} + H(1;1-x) H(-1;-x)\,,
  \end{split}
\end{equation}
and we select the one that gives the smallest imaginary part. This of course give numerically indistinguishable results when the imaginary part is large, but is of
crucial importance when the imaginary part should be zero but it is
not because of the specific numerical methods employed by CHAPLIN.



\bibliographystyle{utcaps}
\bibliography{h1jet.bib}

\end{document}